\documentclass[longauth,linenumbers]{aa}  
\usepackage{graphicx}
\usepackage{xcolor}
\graphicspath{{./}}

\usepackage{amsmath}
\usepackage[hyperfootnotes=false]{hyperref}
\hypersetup{colorlinks=True,linkcolor=black,citecolor=blue}

\usepackage{ulem} %% \sout
\def\blue#1 {{\textcolor{blue}{#1}}\ }
\def\cii{[C\,{\sc ii}]\ }
\def\kms{\,km\,s$^{-1}$}

\def\red#1 {{\textcolor{red}{#1}}\ }
\def\jybeam {{$\text{Jy}\,\text{beam}^{-1}$\ }}

\usepackage{txfonts}
\begin{document}

   \title{The ALMA-CRISTAL survey}

   \subtitle{Widespread dust-obscured star formation in typical star-forming galaxies at $z=4-6$}

   \author{Ikki Mitsuhashi,
          \inst{1,2}
          Ken-ichi Tadaki,
          \inst{3}
          Ryota Ikeda,
          \inst{4,2}
          Rodrigo Herrera-Camus,
          \inst{5}
          Manuel Aravena,
          \inst{6}
          Ilse De Looze,
          \inst{7,8}
          Natascha M. {F{\"o}rster Schreiber},
          \inst{9}
          Jorge {Gonz{\'a}lez-L{\'o}pez},
          \inst{10,6}
          Justin Spilker,
          \inst{11}          
          Roberto J. Assef,
          \inst{6}
          Rychard Bouwens,
          \inst{12}
          Loreto Barcos-Munoz,
          \inst{13,14}
          Jack Birkin,
          \inst{11}
          Rebecca A. A. Bowler,
          \inst{15}
          Gabriela Calistro Rivera,
          \inst{16}
          Rebecca Davies,
          \inst{17,18}
          Elisabete Da Cunha,
          \inst{18,19}
          Tanio {D{\'i}az-Santos},
          \inst{6,20,21}
          Andrea Ferrara,
          \inst{22}
          Deanne Fisher,
          \inst{17}
          Lilian L. Lee,
          \inst{9}
          Juno Li,
          \inst{18,19}
          Dieter Lutz,
          \inst{9}
          Monica {Rela{\~n}o},
          \inst{23,24}
          Thorsten Naab,
          \inst{25}
          Marco Palla,
          \inst{26,7}
          Ana Posses,
          \inst{6}
          Manuel Solimano,
          \inst{6}
          Linda Tacconi,
          \inst{9}
          Hannah {{\"U}bler},
          \inst{27,28}
          Stefan van der Giessen,
          \inst{7,12,25}
          \and
          Sylvain Veilleux
          \inst{29}
          }

   \institute{Department of Astronomy, The University of Tokyo, 7-3-1             Hongo, Bunkyo, Tokyo 113-0033, Japan
              %\email{}
         \and
             National Astronomical Observatory of Japan, 2-21-1 Osawa, Mitaka, Tokyo 181-8588, Japan\\
             \email{ikki0913astr@gmail.com}
         \and
             Faculty of Engineering, Hokkai-Gakuen University, Toyohira-ku, Sapporo 062-8605, Japan
         \and
             Department of Astronomical Science, SOKENDAI (The Graduate University for Advanced Studies), Mitaka, Tokyo 181-8588, Japan
         \and
             Departamento de Astronom\'{i}a, Universidad de Concepci\'{o}n, Barrio Universitario, Concepci\'{o}n, Chile
         \and
             Instituto de Estudios Astrof\'{\i}sicos, Facultad de Ingenier\'{\i}a y Ciencias, Universidad Diego Portales, Av. Ej{\'e}rcito 441, Santiago, Chile
         \and
             Sterrenkundig Observatorium, Ghent University, Krijgslaan 281 - S9, B-9000 Ghent, Belgium
         \and
             Department of Physics \& Astronomy, University College London, Gower Street, London WC1E 6BT, UK
         \and
             Max-Planck-Institut f\"{u}r Extraterrestische Physik (MPE), Giessenbachstr., 85748, Garching, Germany
         \and
             Las Campanas Observatory, Carnegie Institution of Washington, Casilla 601, La Serena, Chile   
         \and
             Department of Physics and Astronomy and George P. and Cynthia Woods Mitchell Institute for Fundamental Physics and Astronomy, Texas A\&M University, College Station, TX, USA
         \and
             Leiden Observatory, Leiden University, NL-2300 RA Leiden, Netherlands
         \and
             Joint ALMA Observatory, Alonso de C\'{o}rdova 3107, Vitacura, Santiago, Chile
         \and
             National Radio Astronomy Observatory, 520 Edgemont Road, Charlottesville, VA 22903, USA
         \and
             Jodrell Bank Centre for Astrophysics, Department of Physics and Astronomy, School of Natural Sciences, The University of Manchester, Manchester, M13 9PL, UK
         \and
             European Southern Observatory (ESO), Karl-Schwarzschild-Stra{\ss}e 2, 85748, Garching bei M{\"u}nchen, Germany
         \and
             Centre for Astrophysics and Supercomputing, Swinburne Univ. of Technology, PO Box 218, Hawthorn, VIC, 3122, Australia
         \and
             ARC Centre of Excellence for All Sky Astrophysics in 3 Dimensions (ASTRO 3D), Australia
         \and
             Research School of Astronomy and Astrophysics, Australian National University, Canberra, ACT 2611, Australia
         \and
             Institute of Astrophysics, Foundation for Research and Technology-Hellas (FORTH), Heraklion, 70013, Greece
         \and
             Chinese Academy of Sciences South America Center for Astronomy (CASSACA), National Astronomical Observatories, CAS, Beijing, 100101, PR China
         \and
             Scuola Normale Superiore, Piazza dei Cavalieri 7, I-50126 Pisa, Italy
         \and
             Dept. Fisica Teorica y del Cosmos, Universidad de Granada, Granada, Spain
         \and
             Instituto Universitario Carlos I de F\'{i}sica Te\'{o}rica y Computacional, Universidad de Granada, E-18071 Granada, Spain        
         \and
             Max-Planck Institute for Astrophysics, Karl Schwarzschildstrasse 1, 85748, Garching, Germany
         \and
             INAF - OAS, Osservatorio di Astrofisica e Scienza dello Spazio di Bologna, via Gobetti 93/3, 40129 Bologna, Italy 
         \and
             Cavendish Laboratory, University of Cambridge, 19 J.J. Thomson Avenue, Cambridge, CB3 0HE, UK
         \and
             Kavli Institute for Cosmology, University of Cambridge, Madingley Road, Cambridge, CB3 0HA, UK 
         \and
             Department of Astronomy and Joint Space-Science Institute, University of Maryland, College Park, Maryland, USA 20742
             }

   \date{Received -; accepted -}

% \abstract{}{}{}{}{} 
% 5 {} token are mandatory
 
  \abstract
   {We present the morphological parameters and global properties of dust-obscured star formation in typical star-forming galaxies at $z=4$--6.
   Among 26 galaxies composed of 20 galaxies observed by the Cycle-8 ALMA Large Program, CRISTAL, and six galaxies from archival data, we have individually detected rest-frame 158$\mu$m dust continuum emission from 19 galaxies, nine of which are reported for the first time.
   The derived far-infrared luminosities are in the range $\log_{10} L_{\rm IR}\,[L_{\odot}]=$10.9--12.4, an order of magnitude lower than previously detected massive dusty star-forming galaxies (DSFGs). 
   The average relationship between the fraction of dust-obscured star formation ($f_{\rm obs}$) and the stellar mass is consistent with previous results at $z=4$--6 in a mass range of $\log_{10} M_{\ast}\,[M_{\odot}]\sim9.5$--11.0 and show potential evolution from $z=6$--9.
   The individual $f_{\rm obs}$ exhibits a significant diversity, and it shows a correlation with the spatial offset between the dust and the UV continuum, suggesting the inhomogeneous dust reddening may cause the source-to-source scatter in $f_{\rm obs}$. 
   The effective radii of the dust emission are on average $\sim$1.5 kpc and are $\sim2$ times more extended than the rest-frame UV. 
   The infrared surface densities of these galaxies ($\Sigma_{\rm IR}\sim2.0\times10^{10}\,L_{\odot}\,{\rm kpc}^{-2}$) are one order of magnitude lower than those of DSFGs that host compact central starbursts. 
   On the basis of the comparable contribution of dust-obscured and dust-unobscured star formation along with their similar spatial extent, we suggest that typical star-forming galaxies at $z=4$--6 form stars throughout the entirety of their disks.}

   \keywords{galaxies:high-redshift --
                galaxies: ISM --
                dust, extinction
               }

   \titlerunning{Resolved dust of normal SFGs at $z=4$--6}
   \authorrunning{Mitsuhashi et al.}
   \maketitle
%
%-------------------------------------------------------------------

\section{Introduction}
In recent decades, star formation activity of galaxies at $z\gtrsim4$ has been progressively studied through their rest-frame ultraviolet (UV) emissions \citep[e.g.,][]{1999ApJ...519....1S,2004ApJ...617..746D,2004ApJ...616L..79B,2007ApJ...670..928B}. 
Because the rest-frame UV continuum emission traces massive young stars that have a short ($\sim$100\, Myr) lifetime, they are used to estimate star formation rates (SFRs) through the empirical relations \citep[e.g.,][]{1998ARA&A..36..189K,2012ARA&A..50..531K}. 
The {\it Hubble Space Telescope} ({\it HST}) provides us with high spatial resolution images covering the rest-frame UV emissions of galaxies at $z\gtrsim4$. 
The spatial extent of the star formation is a fundamental parameter for studying galaxy evolution \citep[e.g.,][]{2003MNRAS.343..978S}, and has been actively explored since the advent of the {\it HST} \citep[e.g.,][]{1998ApJ...500...75L,2004ApJ...611L...1B,2004ApJ...600L.107F,2005ApJ...626..680D,2006ApJ...650...18T,2007ApJ...671..285T}.
%We can understand the morphological evolution of these sizes using radial profile parameters, such as the effective radius assuming a S{\'e}rsic profile \citep{1963BAAA....6...41S}.
%The effective radius of galaxies, which is mostly characterized by the S{\'e}rsic profile \citep{1963BAAA....6...41S}, is an important parameter for understanding morphological evolution.
One of the most surprising results from these studies is that the rest-frame UV effective radii of galaxies decreases with increasing redshift \citep[e.g.,][]{2008AJ....135..156H,2010ApJ...709L..21O,2014ApJ...788...28V,2016ApJ...821...72S}.

The UV emission is highly sensitive to absorption by interstellar dust, and the absorbed energy is re-emitted as thermal dust continuum radiation at far-infrared (FIR) wavelengths.
Therefore, dust continuum emission is an important tracer to study obscured star formation.
In the past decade, Atacama Large Millimeter/Sub-Millimeter Array (ALMA) observations have provided valuable insights into the spatially-resolved properties of high-redshift galaxies, helping us to understand the underlying physical mechanisms driving their morphological evolution.
For instance, submillimeter morphologies of dusty star-forming galaxies (DSFGs) have been well-explored thanks to the high spatial resolution achieved with ALMA 
\citep[e.g.,][]{2015ApJ...810..133I,2016ApJ...833..103H,2017ApJ...850...83F,2019MNRAS.490.4956G}.
DSFGs show intense dust-obscured star formation \citep[][, for review]{2014PhR...541...45C}.
%, and they contribute a significant fraction of the cosmic star formation at $z<3$ \citep{2014MNRAS.438.1267S,2020MNRAS.494.3828D}.
The observed compact star-forming region compared with the stellar distribution \citep[e.g.,][]{2015ApJ...799..194C,2020ApJ...901...74T} suggest that they are in process of the forming bulge \citep[e.g.,][]{2015ApJ...799...81S,2019ApJ...870..130N,2022ApJ...933...11I} or to be progenitor of compact quiescent galaxies \citep[][]{2014ApJ...782...68T,2016ApJ...827L..32B}.
The exploration of DSFGs at $z<3$ is critical as dust obscures the majority of star formation activity within the redshift range of $1<z<3$ \citep{2009A&A...496...57M,2011A&A...528A..35M,2020MNRAS.494.1894M}.

At redshift exceeding $z>3$, dust-obscured star formation may also be significant, \citep[e.g.,][]{2016MNRAS.461.1100R,2020A&A...643A...8G,2023MNRAS.518.6142A} although it has remained uncertain despite extensive efforts \citep[e.g.,][]{2012ApJ...754...83B,2020MNRAS.494.3828D,2021ApJ...909..165Z,2021ApJ...923..215C,2023arXiv230301658F}.
Recent ALMA deep observations have been used to study the dust-obscured star formation activity of galaxies at $z\gtrsim4$, which are originally selected by their bright UV emissions \citep[e.g.,][]{2015Natur.519..327W,2017ApJ...837L..21L,2020ApJ...900....1F,2021MNRAS.508L..58B,2021MNRAS.505.4838L,2021Natur.597..489F,2022ApJ...928...31S}.
Owing to negative $K$-correction and the redshift evolution of the dust temperature ($T_{\rm dust}$) and star formation activities \citep[e.g.,][]{2018A&A...609A..30S,2014ApJS..214...15S}, it is possible to detect dust continuum emission from galaxies at higher redshift in the frequency coverage of ALMA \citep[see][]{2002PhR...369..111B,2014PhR...541...45C}.
The dust continuum detection of relatively low-mass galaxies such as $M_{\ast}\sim10^{10}\,M_{\odot}$ have been increased especially at $z\gtrsim4$ \citep[e.g.,][]{2015Natur.519..327W,2017ApJ...837L..21L,2018MNRAS.481.1631B,2021MNRAS.508L..58B,2021MNRAS.505.4838L,2021Natur.597..489F,2022MNRAS.510.5088B,2022ApJ...928...31S,2022MNRAS.515.3126I,2022ApJ...931..160B,2022MNRAS.512...58F,2022MNRAS.513.3122S}.
\citet{2022MNRAS.515.3126I} found that a significant fraction of the star formation activity in bright Lyman-break galaxies with the stellar mass of $M_{\ast}\sim10^{9.5}\,M_{\odot}$ at $z\sim7$ is obscured by dust \citep[see also][]{2022ApJ...931..160B,2022MNRAS.512...58F,2022MNRAS.513.3122S,2023MNRAS.518.6142A}, consistent with predictions from cosmological galaxy simulations \citep[e.g.,][]{2017MNRAS.470.3006C}.

%Deep ALMA blank-field survey successfully detect the dust emission from galaxies with $M_{\ast}\sim10^{9.5}\,M_{\odot}$ \citep[e.g.,][]{2020ApJ...901...79A} and several studies with stacking analysis revealed average properties of dust emission down to $M_{\ast}\sim10^{10}\,M_{\odot}$ or $L_{\rm IR}\sim10^{11}L_{\odot}$ at $z>1$ \citep[e.g.,][]{2016ApJ...820...83S,2016A&A...587A.122A,2020MNRAS.491.4724F,2020A&A...643A...4F}, individually detecting dust emission from galaxies with $M_{\ast}\sim10^{10}\,M_{\odot}$ remains challenging given that the contribution of dust-obscured star formation is typically smaller in less massive galaxies \citep[e.g.,][]{2014MNRAS.437.1268H,2017ApJ...850..208W}.
%The individual detections of dust emission are still biased towards high stellar mass galaxies \citep[e.g.,][]{2020ApJ...900....1F} even with the aid of gravitational lensing \citep[e.g.,][]{2017A&A...604A.132L,2017A&A...608A.138G,2021ApJ...908..192S} because star formation in more massive galaxies tends to be more obscured by dust \citep[e.g.,][]{2014MNRAS.437.1268H,2017ApJ...850..208W}.

Since the individual detection of the dust continuum from $z\gtrsim4$ galaxies, there have been extensive efforts to understand the role of dust-obscured star formation in the internal structure of UV-bright galaxies.
%, while high-spatial resolution studies are still limited \citep[e.g.,][]{2021A&A...649A..31H,2023MNRAS.524.1775H}.
For instance, \citet{2022MNRAS.510.5088B} revealed the spatial correlation between dust continuum detection and redder colors \citep[see also,][]{2018MNRAS.481.1631B}.
However, spatially-resolved observations of dust emission from the galaxies with $M_{\ast}\sim10^{10}\,M_{\odot}$ remain challenging even with ALMA \citep[e.g.,][]{2020ApJ...900....1F,2020A&A...643A...7G,2022MNRAS.515.3126I,2023MNRAS.524.1775H} and are still limited \citep[e.g.,][]{2021A&A...649A..31H,2023arXiv231108474B,2023arXiv231111493D}.

In this paper, we investigate dust-obscured star formation activity and the spatial extent of dust emission in 26 normal, representative star-forming galaxies at $z=4$--6 by utilizing data from the ALMA Large Program, \cii Resolved ISM in STar-forming galaxies with ALMA (CRISTAL) and some archival data.
With the ancillary data set including deep {\it HST} images, we examine distributions of the dust-obscured and unobscured star-forming regions in the early Universe. 
The paper is organized as follows: Section 2 provides an overview of the CRISTAL survey sample and data products. Section 3 describes the method of size measurements for the ALMA and {\it HST} data, and reports the results of the size measurements and associated physical properties. In Section 4, we discuss the physical origin of the dust distribution and the morphological evolution of the CRISTAL galaxies. A summary of this study is presented in Section 5. Throughout this paper, we assume a flat universe with the cosmological parameters of $\Omega_{\rm M}=0.3$, $\Omega_{\Lambda}=0.7$, $\sigma_{8}=0.8$, and $H_0=70$ \kms ${\rm Mpc}^{-1}$.

%
%
%
% Section 2
%
%
%
\section{Observation and Data}
\subsection{The ALMA CRISTAL survey}\label{subsec:cristalsurvey}
CRISTAL is an ALMA Cycle-8 large program (2021.1.00280.L; PI: Rodrigo Herrera-Camus, see Herrera-Camus et al. in prep). 
CRISTAL aims to spatially resolve the \cii line and rest-frame 158$\mu$m dust continuum emissions of typical star-forming galaxies at $z\sim$4--6. 
The parent sample of CRISTAL is composed of 75 galaxies whose \cii emission is detected in the ALPINE survey (ALMA Large Program to Investigate ${\rm C}^{+}$ at Early Times; 2017.1.00428.L; PI: O. Le F{\`e}vre \citealt{2020A&A...643A...1L}) with the spatial resolution of $\sim1''$. 
For the CRISTAL program, we selected targets based on spectral energy distribution (SED) modeling with {\sc LePhare} \citep{2020ApJS..247...61F,2020A&A...643A...2B}, with the following criteria: (1) the specific SFR is within a factor of 3 from the star formation main sequence at each redshift \citep{2014ApJS..214...15S}, (2) {\it HST} images are available, (3) stellar mass is larger than $\log_{10} M_{\ast}[M_{\odot}]\geq9.5$. 
All galaxies are located in the COSMOS or GOODS-S field and $\sim80\%$ of the galaxies are covered by the ongoing (or planned) {\it JWST} observations such as the COSMOS-Web treasury program \citep{2022arXiv221107865C}, PRIMER program \citep{2021jwst.prop.1837D} or some GTO programs (e.g., ID1286). 
The mass-selected UV-bright galaxies provide a census of the gas, dust, and stars of typical star-forming galaxies on a kiloparsec scale with a combination of high-resolution ALMA and {\it HST}/{\it JWST} observations. 

In addition, we added six galaxies in the COSMOS field \citep[HZ4, HZ7, HZ10, DC818760, DC873756, VC8326, see also][]{2015Natur.522..455C,2020ApJS..247...61F,2023arXiv231108474B,2023arXiv231111493D} referred to as the CRISTAL+ galaxies (corresponding to CRISTAL-20, 21, 22, 23, 24, 25, respectively) because they have ALMA data with similar spatial resolution and sensitivity as the CRISTAL (2018.1.01359.S and 2019.1.01075.S; PI: Manuel Aravena, 2018.1.01605.S; PI: Rodrigo Herrera-Camus, 2019.1.00226.S; PI: Edo Ibar).
Also, we exclude one galaxy (CRISTAL-18) because it does not exhibit clear \cii and dust continuum detections possibly due to misestimation of \cii frequency.
Hence, the CRISTAL extended sample is composed of 24 target fields with available sensitive ALMA \cii plus dust continuum imaging $\lesssim0.3''$.
%Finally, 25 galaxies were originally included in CRISTAL galaxies.
Details of the survey design and observation setup are presented in Herrera-Camus et al. (in prep).

\subsection{Properties of the CRISTAL galaxies}\label{subsec:globsed}
To uniformly estimate galaxy-integrated physical properties of the CRISTAL galaxies including several serendipitously detected sources in both \cii and dust continuum (see Section \ref{subsec:dustdetect}), and confirm the robustness of the physical parameter estimations in the individual SED fitting code, we perform multi-wavelength SED fitting using the {\sc CIGALE} code \citep{2019A&A...622A.103B}. 

We cross-match positions of the CRISTAL galaxies to publicly available photometric catalogs in COSMOS \citep[COSMOS2015 catalog,][]{2016ApJS..224...24L}\footnote[1]{We find there are spatial offsets between the detected position in the COSMOS2015 catalog and the latest the COSMOS catalog \citep{2022ApJS..258...11W}. For our target galaxies, COSMOS2015 catalog shows better identifications.} and GOODS-S fields \citep[ASTRODEEP catalog,][]{2021A&A...649A..22M}. 
All of the CRISTAL galaxies have counterparts within $1''$. 
The model SEDs are generated with stellar population synthesis models of \citet{2003MNRAS.344.1000B} under a \citet{2003PASP..115..763C} initial mass function with a range of $0.1$--$100M_{\odot}$ assuming exponentially declining star formation histories (SFHs).
We adopt \citet{2000ApJ...533..682C} dust attenuation law and {\sc CLOUDY} \citep{1998PASP..110..761F} nebular emission line template.
Details of the fitting and parameter range are summarized in the Appendix.

\begin{figure}[htbp]
\centering
\includegraphics[width=9cm]{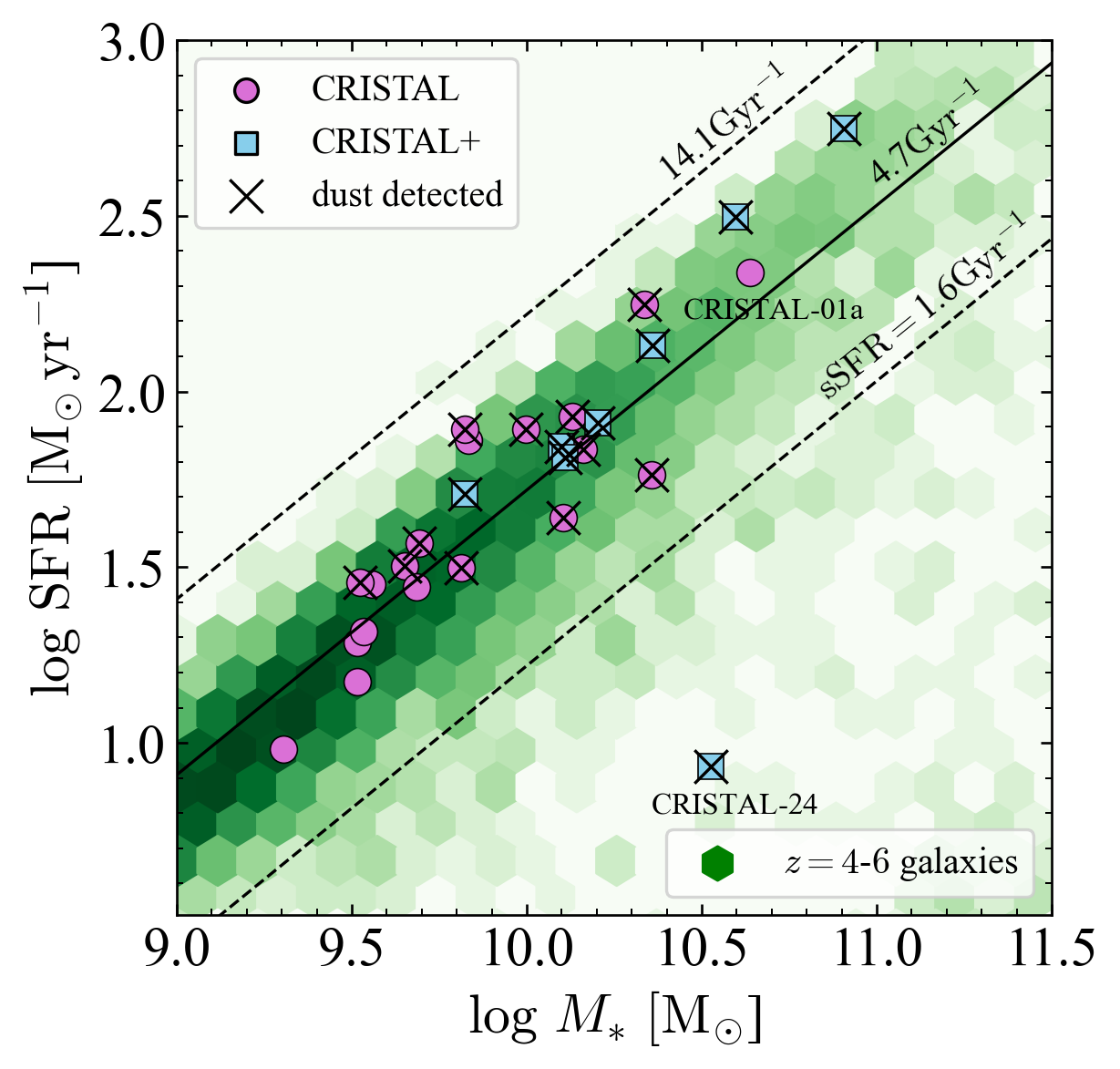}
%\captionsetup{labelformat=empty,labelsep=none}
\caption{SFR vs stellar mass with $z=5$ main-sequence parameterization \citep{2014ApJS..214...15S} and comparison $z=4$--6 galaxies in the COSMOS catalog \citep{2016ApJS..224...24L}. Pink squares are primary targets of the CRISTAL program and blue squares are serendipitously detected galaxies and additional CRISTAL+ galaxies (see Section \ref{subsec:cristalsurvey}). The sources whose dust continuum is detected with ${\rm S/N}>4$ are marked with crosses. The CRISTAL galaxies are located within $\pm0.5$~dex of the star formation main sequence at $z=5$, suggesting they are representative populations at this epoch.}
\label{fig:msplot}
\end{figure}

We derive stellar masses and SFRs averaging the SFH over the last 100 Myr (${\rm SFR}_{\rm SED}$) from the best-fit SED models. 
Figure \ref{fig:msplot} shows the relation between stellar mass and SFR for the CRISTAL galaxies.
We also perform SED fitting for all galaxies at $z=4-6$ in the COSMOS2015 catalog with the exact same setup as for the CRISTAL galaxies.
CRISTAL galaxies have stellar masses of $\log_{10} M_{\ast}[M_{\odot}]\gtrsim9.5$, which are consistent with the results from {\sc LePhare} code in the ALPINE program within the 1$\sigma$ uncertainty.
%, except for one.
A comparison with the star formation main sequence at $z=5$ \citep{2014ApJS..214...15S} indicates that CRISTAL galaxies are a representative population of star-forming galaxies at this epoch.
Note that the estimated parameters of CRISTAL-01 might be uncertain due to the contamination of {\it Spitzer} band photometry by a nearby bright source (see Figure \ref{fig:msplot}).
The stellar mass estimation will be improved by {\it JWST} data in the future (Li et al. in prep) as most of the CRISTAL galaxies have {\it JWST}/NIRCam coverage as described in Section \ref{subsec:cristalsurvey}.
We also note that CRISTAL-24 is located below the main sequence due to small ${\rm SFR}_{\rm SED}$.
Since SFR estimated from UV and IR radiation (${\rm SFR}_{\rm UV}+{\rm SFR}_{\rm IR}$) is $\sim1.3\,{\rm dex}$ higher than ${\rm SFR}_{\rm SED}$, ${\rm SFR}_{\rm SED}$ derived from optical to NIR photometry appears to be underestimated due to the degeneracy of dust extinction and stellar age (see Figure \ref{fig:sfruvir}).
The stellar masses and SFRs for the CRISTAL galaxies are summarized in Herrera-Camus et al. (in prep).

%
%
%
% Figure 1
%
%
%

\subsection{Observation and Data products}

CRISTAL observations cover a wide range of spatial frequencies in the $uv$ plane with a combination of extended (C43-4/5/6) and compact (C43-1/2) array configurations to achieve not only high-resolution ($\sim0.3''$, corresponding to $\sim1.9\,{\rm kpc}$ in physical scale at $z=5$) but also a large maximum recoverable scale ($\sim4.5''$, corresponding to $\sim28.3\,{\rm kpc}$ in physical scale at $z=5$).
We determine the spectral frequency coverages by centering \cii frequency derived from the integrated \cii spectral line profile in the ALPINE survey.
Observations were carried out from November 2021 to April 2023.
Details of the calibration, reduction, and data products will be presented in Herrera-Camus et al. (in prep). 

In addition to the CRISTAL observations taken in two array configurations, all CRISTAL galaxies have archival observations (2012.1.00978.S, 2012.1.00523.S, and 2017.1.00428.L). 
As these archival data typically cover shorter spatial frequency ranges in the $uv$ plane compared with the CRISTAL observations, adding these archival data improves the sensitivities, especially to extended emissions. 
We use the archive data by combining visibilities with the {\sc CASA/concat} task.

%\textbf{products from JGL?}:

%
%
%
% Section 3
%
%
%
\section{Analysis and Results}\label{sec:analysis}
\subsection{Dust continuum detected galaxies}\label{subsec:dustdetect}
In this paper, we focus on the dust continuum emission of CRISTAL galaxies. 
We exclude visibility data in the frequency range corresponding to the \cii emission line. 
To ensure avoiding the contamination of \cii lines to the dust continuum, we exclude $\pm0.45\,{\rm GHz}$, which correspond to $\pm400\,{\rm km}\,{\rm s}^{-1}$ and $\pm450\,{\rm km}\,{\rm s}^{-1}$ for $z=4.5$ and 5.5 respectively, from the \cii central frequencies since broad component of \cii emission lines could reach $\sim250\,{\rm km}\,{\rm s}^{-1}$ in FWHM/2 \citep{2020A&A...633A..90G}.
Next, we make multi-frequency synthesis (continuum) images with {\sc CASA} task {\sc tclean} to look at their signal-to-noise ratio (S/N). 
Here we did not apply {\sc clean} algorithm (i.e., dirty images) and primary beam correction to perform a consistent comparison between noises and signals \citep[c.f., JvM correction,][Gonzalez-Lopez et al. in prep]{2021ApJS..257....2C}.
To avoid low S/N for extended emission in galaxies and investigate how well the extended emission of galaxies is recovered, we create low-resolution images with $uv$ taper of $0.25''$, $0.5''$, $1.0''$, and $2.0''$ as well as high-resolution images with natural weighting. 
In all images with different resolutions, we pick maximum fluxes in dust continuum images within a circle of radius $0.5''$ around the position of previous \cii detection.
%to take a maximum value.
The noise is estimated by calculating 3$\sigma$ clipped root-mean-square (RMS) values of all pixels within the Field-of-View (FoV) where the primary beam correction value is between 0.9 and 0.5 to avoid the impact of the target galaxy for sure.
%Since the number of pixels to calculate the RMS value is much smaller than that of all pixels within the FoV, a spatial correlation of the noise is negligible.
Typical RMSs are 10--20~$\mu$\jybeam.

In the vicinity of the main targets, we serendipitously found 2 sources with $>8\sigma$ \cii detection, $>4\sigma$ dust continuum detection, and {\it HST} detection (CRISTAL-01b and 07c). 
We include these galaxies in the main analysis. 
Please refer to Hererra-Camus et al. (in prep) for more detail on serendipitous \cii and dust detections around CRISTAL galaxies.
In a total of 26 extended CRISTAL samples (18 original CRISTAL targets, 6 CRISTAL+ galaxies, and 2 serendipitously detected galaxies in original CRISTAL fields), we found that 7, 8, and 11 galaxies have dust continuum emissions with ${\rm S/N}<4$, $4\leq{\rm S/N}<6.5$, and ${\rm S/N}\geq6.5$ respectively. 

\begin{figure*}[htbp]
\begin{center}
%%\epsscale{1.15}
\includegraphics[width=16cm]{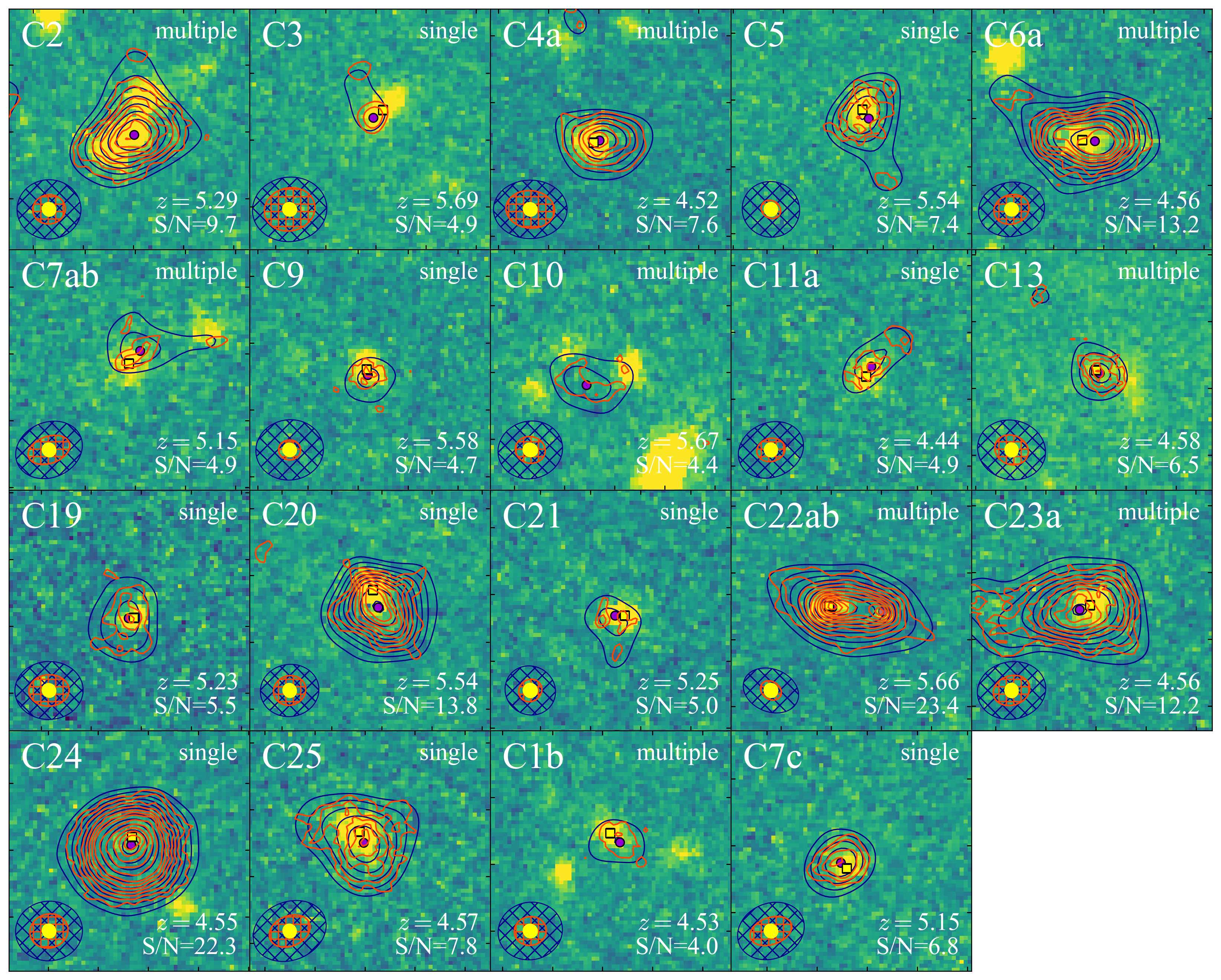}
%\captionsetup{labelformat=empty,labelsep=none}
\caption{Thumbnails of the dust detected CRISTAL galaxies. The background images are {\it HST}/F160W images. The natural-weighted and natural-weighted+$0.5''$-Tapered rest-frame 158 $\mu{\rm m}$ dust continuums are also shown in red and blue contours, respectively. An average beam size of natural-weighted continuum images is $0.46''=2.9\,{\rm kpc}$. The contour levels are shown every 1$\sigma$ within 3--9$\sigma$ and every 2$\sigma$ from 9$\sigma$, respectively. For CRISTAL-22, for every 2$\sigma$ from 3$\sigma$. The source ID, redshift, S/N, and morphological classification based on the {\it HST} images are also shown in each panel.}
\label{fig:thumbneils}
\end{center}
\end{figure*}

We show thumbnails of the dust-detected CRISTAL galaxies in Figure \ref{fig:thumbneils}.
The dust-detected galaxies tend to have higher stellar masses ($\log_{10} M_{\ast}\,[M_{\odot}]\gtrsim9.7$) and SFRs ($\log_{10} {\rm SFR}\,[M_{\odot}\,{\rm yr}^{-1}]\gtrsim1.5$).
The median stellar mass and SFR of the dust-detected and undetected sample are $\log_{10} M_{\ast}\,[M_{\odot}]=10.1$, $\log_{10} {\rm SFR}\,[M_{\odot}\,{\rm yr}^{-1}]=1.8$ and $\log_{10} M_{\ast}\,[M_{\odot}]=9.5$, $\log_{10} {\rm SFR}\,[M_{\odot}\,{\rm yr}^{-1}]=1.4$, respectively.
The detection rate is 89\% (16/18) in a stellar mass range of $\log_{10} M_{\ast}\,[M_{\odot}]>9.7$ (Figure \ref{fig:msplot}).
%while the dust emission is not detected in a lower stellar mass range.
%Dust-obscured star formation activities on massive galaxies with $M_{\ast}\gtrsim10^{10.5}\,M_{\odot}$ are well studied by ALMA observations \citep[e.g.,][]{2016ApJ...827L..32B,2016ApJ...833...12R,2017ApJ...850...83F,2018A&A...616A.110E,2020ApJ...901...74T,2020PASJ...72...69Y}.
%Given that the contribution of dust-obscured star formation is typically smaller in less massive galaxies \citep[e.g.,][]{2014MNRAS.437.1268H,2017ApJ...850..208W}, individually detecting dust emission from galaxies with $M_{\ast}\sim10^{10}\,M_{\odot}$ remains challenging.
%Several studies with stacking analysis revealed average properties of dust emission down to $M_{\ast}\sim10^{10}\,M_{\odot}$ or $L_{\rm IR}\sim10^{11}L_{\odot}$ at $z>1$ \citep[e.g.,][]{2016ApJ...820...83S,2016A&A...587A.122A,2020MNRAS.491.4724F,2020A&A...643A...4F}. 
%One successful example of the detection of dust emission from galaxies with $M_{\ast}\sim10^{9.5}\,M_{\odot}$ is reported by deep ALMA blank-field survey \citep[e.g.,][]{2020ApJ...901...79A}. 
%Some attempts to detect faint dust emission utilizing the gravitational lensing effect have been also conducted in less massive galaxies \citep{2017A&A...608A.138G,2017A&A...604A.132L,2021ApJ...908..192S}.
Our detections of faint galaxies in the dust continuum down to the stellar mass of $M_{\ast}\sim10^{9.5}\,M_{\odot}$ provide new insight into the dust-obscured star formation activities of less massive galaxies.

%
%
%
% Figure 2
%
%
%

\subsection{Dust size measurement}\label{subsec:dustsize}
%\subsubsection{Dust size and flux measurement}
In this section, we characterize the spatial extent of the rest-frame 158$\mu {\rm m}$ dust continuum emissions in the CRISTAL galaxies.
In principle, it is possible to measure the size of galaxies by deconvolving an image with a clean beam. 
However, reconstructed images through the Fourier transform of the visibility data and non-linear {\sc clean} algorithm strongly depend on the $uv$ coverage of visibilities and its weight, which could lead to systematic uncertainties. 
Moreover, as about one-third of CRISTAL galaxies lack high S/N (${\rm S/N}\gtrsim5$) in the dust continuum images, image-based analysis may make measurements uncertain.
To achieve homogenous size measurements and to utilize full information from the observations, we directly measure the effective radii of dust continuum emission from the visibility data. 

%The procedure for our size measurement by visibility fitting is as follows.
Recent high-resolution ALMA observations revealed that DSFGs are generally consistent with S{\'e}rsic profile \citep{1963BAAA....6...41S} with the index of $n\sim1$ \citep[e.g.,][]{2016ApJ...833..103H,2019MNRAS.490.4956G}. 
Even for fainter sources, \citet{2020ApJ...901...74T} suggest that their surface density profiles are better characterized by an exponential profile with $n=1$ than a Gaussian profile with $n=0.5$. Therefore we fit a 2D exponential profile ($n=1$) by using {\sc UVMultifit} \citep{2014A&A...563A.136M}.

The observed visibilities are presented as a combination of amplitude and phase from all sources within the FoV. 
Hence, to extract information specific to the target galaxy, it is necessary to subtract contributions from other continuum sources within the FoV especially when they are close to the target.
Sources more than $2''$ apart from the target galaxy are subtracted using the {\sc CASA} {\sc tclean} task.
We clean down to 1.5$\sigma$ by applying $2''$ masks from the source position to extract the clean component of the nearby sources and subtract them from the observed visibilities with {\sc CASA} task {\sc uvsub}.
Note that there is no significant difference if we conduct the two-component fitting to take into account the other continuum sources.
When we clearly find the existence of a second component in the residual map after subtracting the best-fit single component with ${\rm S/N}>10$ (CRISTAL-22), we use the results of the two-component fitting.
As the CRISTAL galaxies typically do not have high S/N (${\rm S/N}>5$) in the dust emission (Table \ref{tab:tab1}), the axis ratio ($q$) is not well constrained in most galaxies. 
Thus, we measure the effective radii of the dust emission by assuming a circular exponential disk with a fixed value of $q=0$.
We create residual maps from the visibilities where the best-fit model is subtracted in the $uv$ plane. 

\begin{figure*}[htbp]
\begin{center}
%\epsscale{1.15}
\vspace{1cm}
\includegraphics[width=17cm]{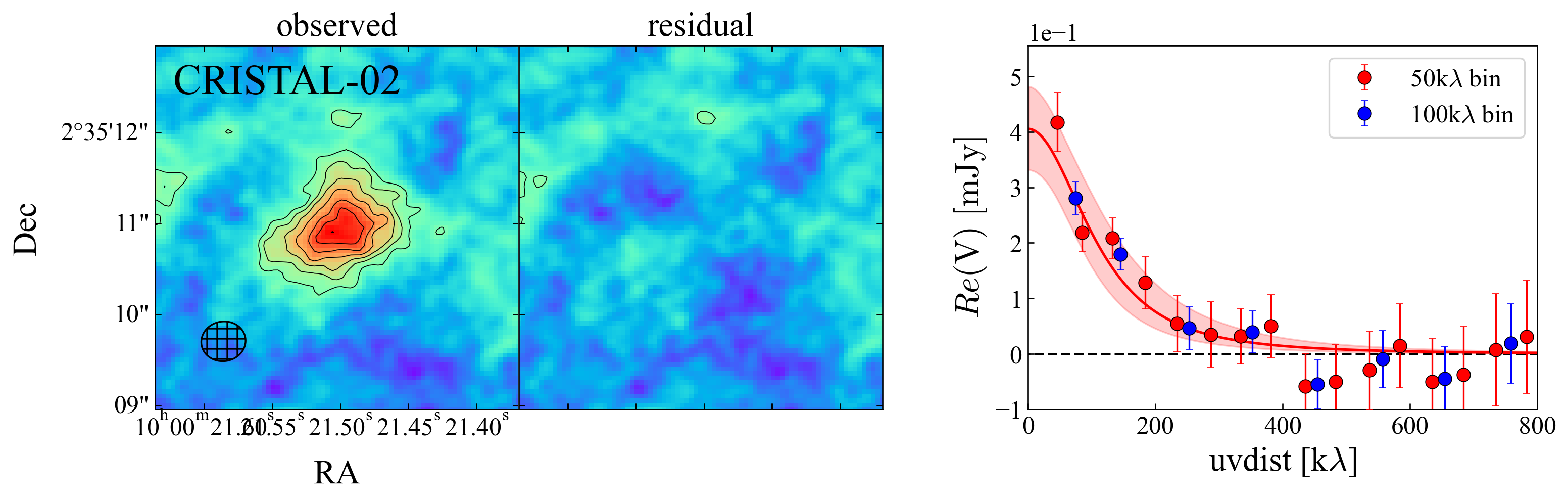}
%\captionsetup{labelformat=empty,labelsep=none}
\caption{An example of the fitting in the visibility plane. {\bf left:} The observed and residual (observed - best-fit model) dirty map of the dust continuum emission. The synthesized beam is shown in the bottom left. The black contour shows every 1$\sigma$ from $\pm3\sigma$. {\bf right:} The real part of the observed visibilities as a function of $uv$ distances with different binning scales. The best-fit exponential disk model and its 1$\sigma$ uncertainty are shown in a red solid line and shaded area respectively.}
\label{fig:fiteg}
\end{center}
\end{figure*}

Figure \ref{fig:fiteg} shows an example of a dirty map, a residual map, and a real part of the observed visibilities as a function of $uv$ distance.
Because there is no peak above $\sim3\sigma$ in the residual maps, the exponential disk model seems to successfully represent the spatial distribution of dust emission. 
Thanks to the observations in the extended array configurations, we successfully capture a decline of the real part of the visibilities that clearly suggests the dust continuum from the source is spatially resolved.
A signal at the real part with increasing $uv$ distance is the reciprocal of the spatial extent at an image plane since the Fourier transform of the profile in the image plane is a profile in the visibility plane.
All of the original, model, and residual maps are shown in the Appendix.

%
%
%
% Figure 3
%
%
%

\begin{figure}[t]
\centering
\includegraphics[width=9cm]{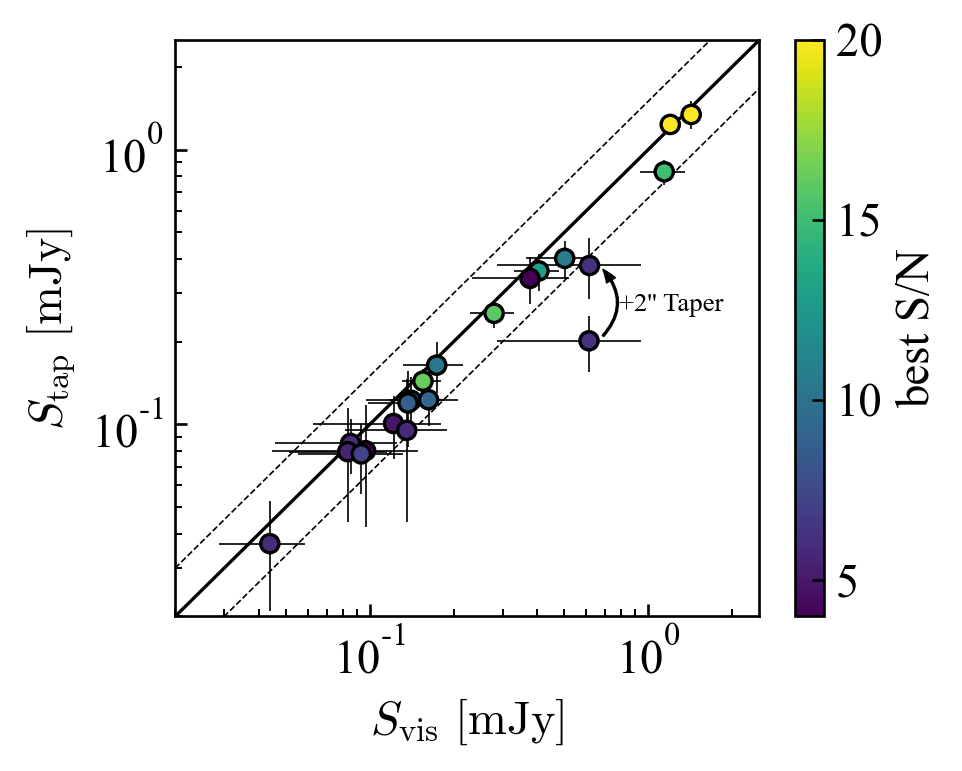}
%\captionsetup{labelformat=empty,labelsep=none}
\caption{Comparison of the flux density estimated from {\sc UVMultifit} and Taper images, color-coded by the source S/N. The black solid line and dashed lines indicate a one-to-one relation and $\pm0.3\,{\rm dex}$. We see the visibility fitting recovered the fluxes of the sources in lower resolution images well within the 1-$\sigma$ errors, except for CRISTAL-21 (see text).}
\label{fig:fluxcomp}
\end{figure}

To check the goodness of the visibility fitting, we compare the flux densities of dust continuum emission estimated from the visibility fitting and those from the low-resolution image (Figure \ref{fig:fluxcomp}). 
Here we make the images with a beam size of $3.0''$ by applying {\sc CASA/imsmooth} to the $2.0''$-tapered images to transform the resolution of each source to a common beam size and to make sure that all of the flux densities are included within a single beam. 
Then we measured source flux densities with the same method used in Section \ref{subsec:dustdetect}. 
%Note that if we use the images with different taper scales, the results are not significantly changed. 
We find the visibility fitting recovers the flux densities well in various sources from the brightest to the faintest among the CRISTAL galaxies, except for a very extended source (CRISTAL-21). 
The inconsistency of flux densities in CRISTAL-21 likely originates from extrapolation of the best-fit model toward the short $uv$ distance regime due to a low sensitivity at $\leq50~{\rm k}\lambda$.
When the additional $2''$-taper is applied to CRISTAL-21, we find a better match with the visibility-based estimation.
To maintain consistency between the detected and undetected galaxies in the dust continuum, hereafter we use the flux densities and 1$\sigma$ uncertainties estimated from the low-resolution (i.e., tapered) images.

We summarize the fluxes measured in visibilities via {\sc UVMultifit} and low-resolution images in Table \ref{tab:tab1}.
Since about half of the CRISTAL galaxies range $4.4<{\rm S/N}<5.5$ in the dust continuum images (see Table \ref{tab:tab1}), we conduct a Monte Carlo simulation and a stacking analysis to confirm the validity of the size measurement in Appendix. 
We do not find any systematic offset regardless of the source S/N in the Monte Carlo simulation and there is no clear difference between the stacked size and the average of the individual sizes in the stacking analysis.
Therefore we conclude that the fitting results are reliable.

%
%
%
% Figure 4
%
%
%

\subsection{FIR luminosity}
Because most of the CRISTAL galaxies have only single band ALMA detections in the dust continuum, it is not possible to estimate properties such as dust temperature ($T_{\rm dust}$) and emissivity ($\beta_{\rm dust}$), which are crucial in determining the shape of the FIR SED. 
\citet{2017A&A...607A..89B} presented best-fit \citet{2007ApJ...657..810D} dust emission model for stacked fluxes of the main-sequence galaxies at $z\sim4$ with {\it Spitzer}, {\it Herschel}, LABOCA, and AzTEC data \citep[see also][]{2015A&A...573A.113B}. 
This best-fit FIR SED agrees with the stacked fluxes of $z=4$--6 main-sequence galaxies with ${\rm SFR}_{\rm SED}>10\,M_{\odot}\,{\rm yr}^{-1}$ \citep{2020A&A...643A...2B}, corresponding to the modified blackbody with the dust temperature $T_{\rm dust}\sim42\,{\rm K}$. 
Indeed, some CRISTAL galaxies have rest-frame 110 and 200 $\mu$m observations and show a consistent $T_{\rm dust}$ estimation with the assumption of $T_{\rm dust}\sim42\,{\rm K}$ \citep{2020MNRAS.498.4192F}, while there exists a variation in the FIR SED with the range of dust temperatures of $T_d\sim30$--60\,K.
\citet{2022MNRAS.517.5930S} apply the physically-motivated model to characterize $T_{\rm dust}$ validated in \citet{2021MNRAS.503.4878S} for the $z\sim5$ galaxies that is partly overlapping with the CRISTAL galaxies.
The derived average $T_{\rm dust}$ is $48\pm8\,{\rm K}$, and is broadly consistent with the assumption of $42\,{\rm K}$.
Because our CRISTAL galaxies share similar SFRs with the galaxies studied in \citet{2020A&A...643A...2B}, we use the best-fit FIR SED presented in \citet{2017A&A...607A..89B} to estimate the total IR (8--$1000\,\mu{\rm m}$) luminosities of the CRISTAL galaxies. 
We calculate the total FIR luminosities of the CRISTAL galaxies by converting the measured dust continuum fluxes at the rest-frame 158\,$\mu$m using a scaling factor described \citet[][$\nu L_{\nu=158\mu{\rm m}}/L_{\rm IR}=0.133$]{2020A&A...643A...2B}.
%We convert measured dust continuum fluxes at rest-frame 158\,$\mu$m to total IR luminosities by applying a scaling factor presented in \citet[][$\nu L_{\nu=158\mu{\rm m}}/L_{\rm IR}=0.133$]{2020A&A...643A...2B}.
The resulting $L_{\rm IR}$ of CRISTAL galaxies fall within a range of $\log_{10}L_{\rm IR}\,[L_{\odot}]=10.9$--12.4.
%and are nearly comparable to that of the local (Ultra-) luminous infrared galaxies \citep[(U)LIRGs, e.g.,][]{2017ApJ...846...32D}.

\begin{figure*}[htbp]
\centering
\includegraphics[width=18cm]{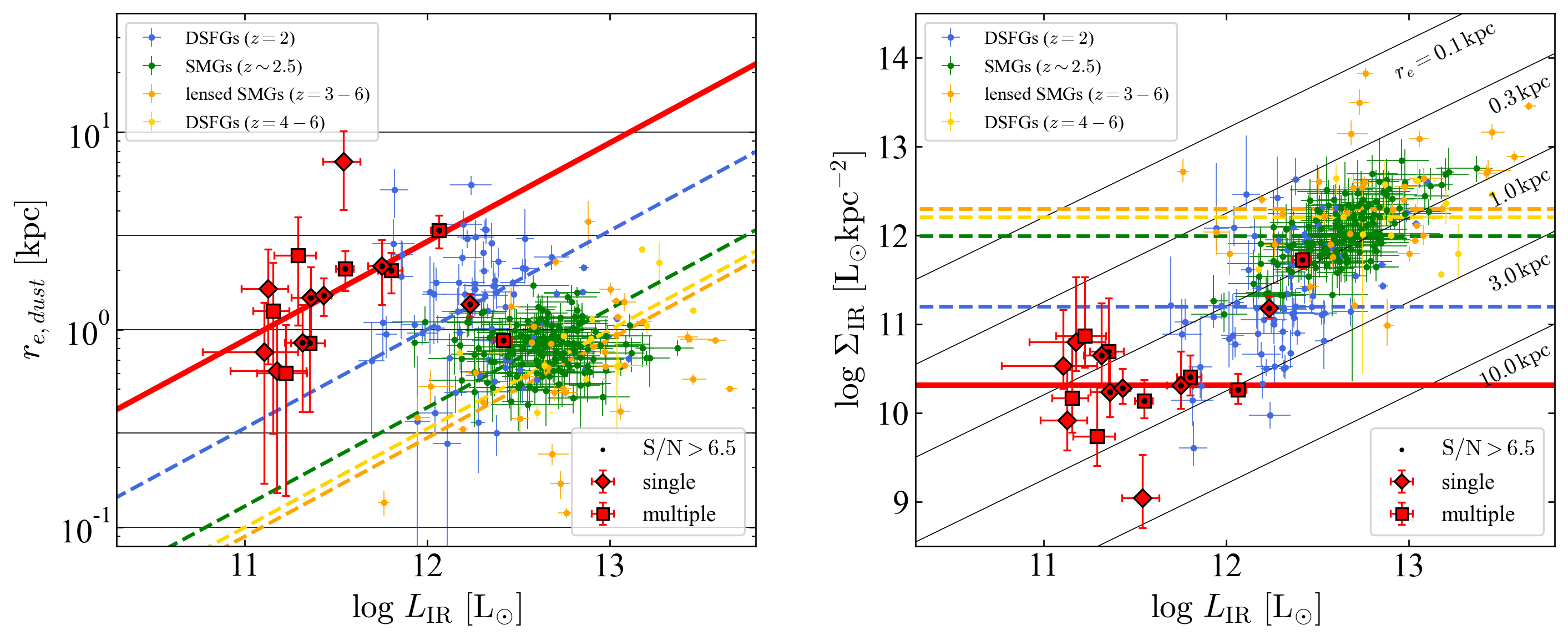}
%\captionsetup{labelformat=empty,labelsep=none}
\caption{Total infrared luminosities and {\bf (left)} effective radii and {\bf (right)} surface infrared lumnosities for CRISTAL galaxies, SMGs at $z\sim2.5$ \citep{2019MNRAS.490.4956G}, lensed SMGs at $z\sim3$--6 \citep{2016ApJ...826..112S}, SMGs at $z=4$--6 \citep{2014ApJ...796...84R,2014A&A...565A..59D,2015ApJ...798L..18H,2018ApJ...856..121G,2019ApJ...887...55C,2020ApJ...889..141T} and DSFGs at $z\sim2$ \citep{2020ApJ...901...74T}. Solid and dashed lines show the median $\Sigma_{\rm IR}$ for each sample. CRISTAL galaxies show $\sim$10 times smaller $\Sigma_{\rm IR}$ than that of DSFGs or SMGs, and are less extreme galaxies compared with DSFGs.}
\label{fig:LvsR}
\end{figure*}

Figure \ref{fig:LvsR} shows the total infrared luminosity against the circularized effective radius (left) or the surface infrared luminosity $\equiv\Sigma_{\rm IR}=L_{\rm IR}/2\pi r_{e, {\rm dust}}^2$ (right) of dust continuum emission. 
As a comparison, we show the results of different galaxy populations, namely sub-millimeter galaxies (SMGs) at $z\sim2.5$ \citep{2019MNRAS.490.4956G}, lensed SMGs at $z\sim3$--6 \citep{2016ApJ...826..112S}, DSFGs at $z=4$--6 \citep{2014ApJ...796...84R,2014A&A...565A..59D,2015ApJ...798L..18H,2018ApJ...856..121G,2019ApJ...887...55C,2020ApJ...889..141T} and DSFGs at $z\sim2$ \citep{2020ApJ...901...74T}.
The comparison galaxies are mainly starburst galaxies above the main sequence at the corresponding epochs, and their FIR luminosities are derived by some templates of FIR SEDs \citep[e.g.,][]{2008ApJ...682..985W,2015ApJ...806..110D} or modified blackbody fitting with the typical dust temperature $T_{\rm dust}=30$--$50\,{\rm K}$. 
Solid and dashed lines indicate a series of $r_e$ and $L_{\rm IR}$ each corresponding to the median infrared surface densities ($\Sigma_{\rm IR}$) of the respective sample, as indicated by the matching color.

The median values of $\Sigma_{\rm IR}$ are $2.0\times10^{12}\,L_{\odot}\,{\rm kpc}^{-2}$ for lensed SMGs, $9.8\times10^{11}\,L_{\odot}\,{\rm kpc}^{-2}$ for SMGs at $z\sim2.5$, $1.6\times10^{12}\,L_{\odot}\,{\rm kpc}^{-2}$ for SMGs at $z=4$--6, and $1.6\times10^{11}\,L_{\odot}\,{\rm kpc}^{-2}$ for DSFGs at $z=2$.
The median $\Sigma_{\rm IR}$ of the CRISTAL galaxies is $2.0\times10^{10}\,L_{\odot}\,{\rm kpc}^{-2}$ and $\sim$10 times smaller than that of the central compact star-forming regions of DSFGs or SMGs (Figure \ref{fig:LvsR} right), and similar to that of the extended dust continuum component of the SMGs reported in the stacked spatial distribution \citep[][$\Sigma_{\rm IR}\sim5\times10^{10}\,L_{\odot}\,{\rm kpc}^{-2}$]{2019MNRAS.490.4956G} that is considered to be disk components.
Our representative population of galaxies at $z\sim$4--6 seems to undergo moderate star formation with a lower $\Sigma_{\rm IR}$ than those of DSFGs, except for some galaxies (CRISTAL-22 and 24) showing the high $\Sigma_{\rm IR}$ those are comparable with DSFGs and SMGs.
Interestingly, \citet{2022MNRAS.512...58F} analytically derive the $L_{\rm IR}$ and dust distribution radius of the REBELS \citep[Reionization Era Bright Emission Line Survey;][]{2022ApJ...931..160B} galaxies at $z\sim7$.
They find a more compact distribution ($r_d\sim0.3\,{\rm kpc}$) and a similar $L_{\rm IR}$ with the CRISTAL galaxies ($\log_{10} L_{\rm IR}\,[L_{\odot}]\sim11.8$), which results in a higher $\Sigma_{\rm IR}$ ($\log_{10}\Sigma_{\rm IR}\,[L_{\odot}\,{\rm kpc}^{-2}]\sim12.1$) than our measurements.
Direct measurement of the dust continuum distribution by the future ALMA high-resolution observations for $z>6$ galaxies will enable us to compare our results with that of the galaxies at $z>6$.

%
%
%
% Figure 6
%
%
%

\subsection{Rest-frame UV size measurement}\label{subsec:UVsize}
To compare the spatial extent of the dust-obscured and unobscured star formation, we measure the rest-frame UV sizes of the CRISTAL galaxies. In this section, we summarize the size measurements of the rest-frame UV emission observed with {\it HST}.

Because all of the CRISTAL galaxies have WFC3/F160W coverage and 1.6\,$\mu$m corresponds to rest-frame $\sim$2400--3000\,\AA\ in the redshifts of the CRISTAL galaxies ($z=4.4$--5.7), we constantly use images from WFC3/F160W observation to measure the rest-frame UV sizes of the CRISTAL galaxies.

We measure the {\it HST} size based on two-dimensional (2D) surface-brightness profile fitting with {\sc GALFIT} \citep{2002AJ....124..266P}. 
First, we make a point-spread function (PSF) in the same manner as previous studies \citep[e.g.,][]{2014ApJS..214...24S}. 
We selected point sources (i.e., stars) based on the following criteria: (1) ratios of fluxes with $2.0''$ apertures and $0.5''$ apertures are in a range of 1.2 and 1.4 based on the tight flux correlation in the point source. (2) S/Ns are greater than 7500.
We exclude the point sources if they have nearby sources within $3''$. 
Because the central position of the point source does not necessarily correspond to the central position of the pixel, we reproject the 2D point source profile with a linear interpolation to the central position derived from 2D Gaussian fitting.
Then we obtain the PSF by stacking these reprojected stellar profiles after normalizing with the total flux of each source.
We also create {\it sigma} and {\it mask} images to estimate the weight of individual pixels and exclude neighboring objects of the target galaxy from the fittings. 
The {\it sigma} images are made from drizzle weight maps and {\it mask} images are produced from segmentation maps from {\sc SExtractor} with a parameters of {\sc detect\_minarea}=5, {\sc detect\_thresh}=2, {\sc detect\_nthresh}=16 and {\sc deblend\_mincont}=0.0001.
We input the {\it sigma}, {\it mask}, and PSF to {\sc GALFIT} and fit a single S{\'e}rsic profile to the 2D surface-brightness profile of each galaxy. 
Here we fix $n=1$ to perform fair comparisons with the dust size measurement. 
Note that there is no systematic difference if we do not fix the S{\'e}rsic index except for CRISTAL-25, showing a large S{\'e}rsic index in the best-fit model (see Figure \ref{fig:sizecomp}).
Finally we obtain an effective radius along the major axis $r_{e, {\rm UV}}^{\rm major}$ and axis ratio $q$, and convert them to the circularized effective radius $r_{e, {\rm UV}}$ through $r_{e, {\rm UV}}\equiv r_{e, {\rm UV}}^{\rm major}\sqrt{q}$.
If the fitting does not converge with the axis ratio as a free parameter, we fix the axis ratio to unity as in the size measurement in the dust continuum.
The original, model, and residual maps are shown in Appendix.
For CRISTAL-22, we only measure the size of the main component by masking the second component because the latter is too faint and the fitting does not converge (see Figure \ref{fig:thumbneils}).

\begin{figure}[htbp]
\centering
\includegraphics[width=8.5cm]{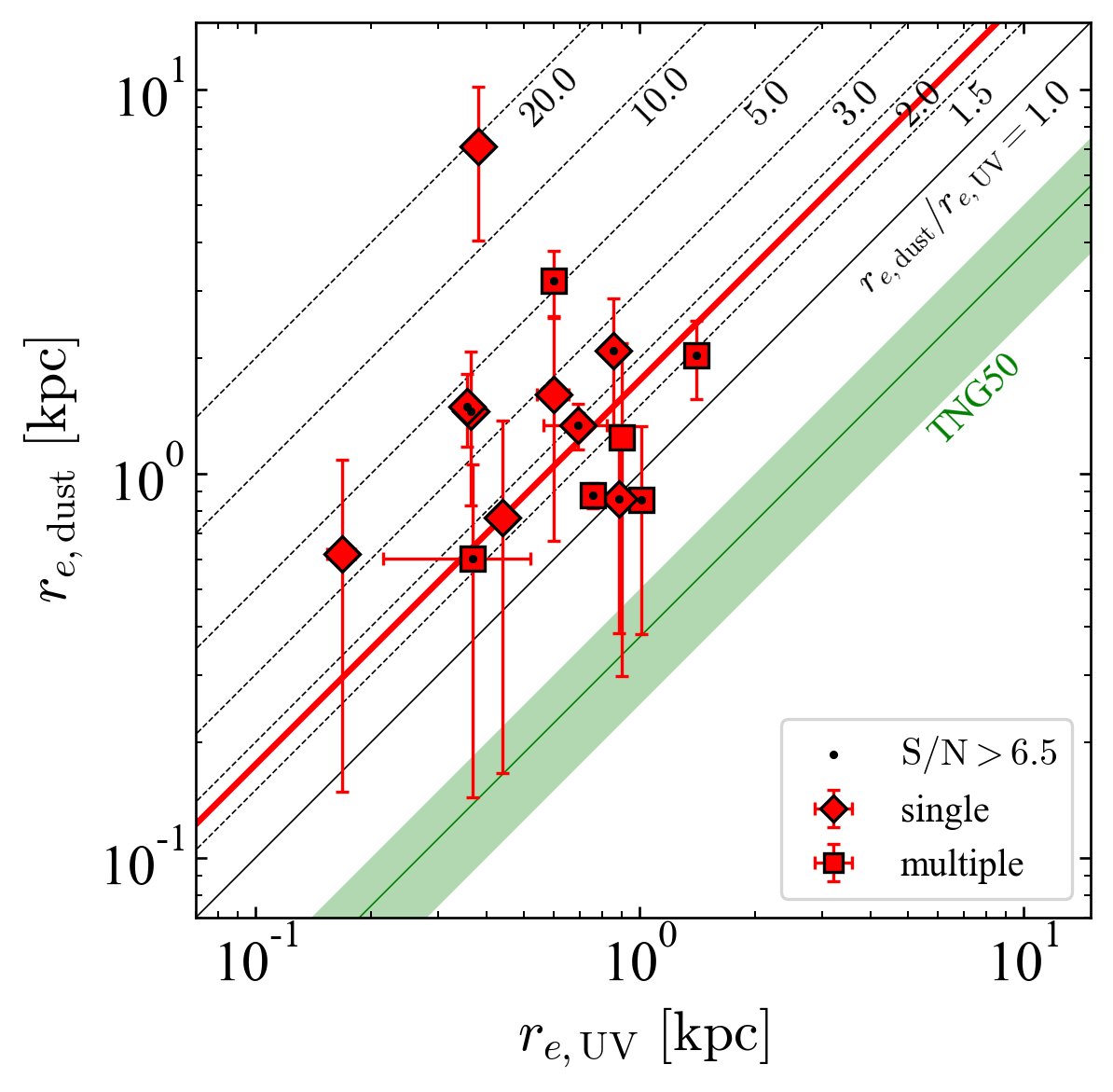}
\caption{Size comparison of the dust continuum ($r_{e,{\rm dust}}$) and the UV continuum ($r_{e,{\rm UV}}$). The black solid and dashed lines indicate the size ratios of the UV and dust continuum. An average size ratio of the CRISTAL galaxies ($r_{e,{\rm dust}}/r_{e,{\rm UV}}\sim1.8$) are shown in the red solid line. The green shaded region shows the size ratio predicted in illustrious TNG50 simulation \citep{2022MNRAS.510.3321P}. Overall, CRISTAL galaxies possibly have a more extended distribution in dust thermal emission than in UV radiation, and it is contrary to the expectation from the simulation.}
\label{fig:reuvreir}
\end{figure}

In Figure \ref{fig:reuvreir}, we compare measured sizes of the dust-obscured star formation derived from rest-frame 158\,$\mu$m dust continuum observations by ALMA and the dust-unobscured star formation derived from rest-frame UV continuum observations by {\it HST}/WFC3 F160W. 
The dust-obscured star formation appears to be on average $\sim2$ times more spatially extended than the dust-unobscured star formation. 
We discuss the spatial extent of the dust and UV emissions in Sec \ref{subsubsec:exdust}.

The measured dust and rest-frame UV sizes are summarized in Table \ref{tab:tab1}.

\renewcommand{\arraystretch}{1.2}
\tabcolsep = 0.1cm
%
%
%
% Table 1
%
%
%
\begin{table*}
%\tablenum{1}
\centering
\caption{Dust continuum properties of the CRISTAL galaxies}
\begin{tabular}{lccccccccc}
%\tablewidth{0pt}
\hline
ID & $\log_{10} M_{\ast}$ & ${\rm S/N}$ & 
$S_{\rm uv}$\tablefootmark{\small a} & $S_{\rm tap}$\tablefootmark{\small b} & $\log_{10} L_{\rm IR}$ & $f_{\rm obs}$ & $r_{e,{\rm dust}}$ & $r_{e,{\rm UV}}$\tablefootmark{\small c} & class\tablefootmark{\small d}\\
 & [$M_{\odot}$] &  & ${\rm [mJy]}$ & ${\rm [mJy]}$ & [$L_{\odot}$] &  & [arcsec] & [arcsec] & \\
\hline
\multicolumn{9}{c}{{\bf CRISTAL main sample}} \\
CRISTAL-01a & $10.65\pm0.50$ & - & - & $\leq0.251$ & $\leq11.54$ & $\leq0.426$ & - & $0.154\pm0.007$ & multiple\\
CRISTAL-02 & $10.30\pm0.28$ & 9.3 & $0.405\pm0.075$ & $0.362\pm0.055$ & $11.80_{-0.07}^{+0.06}$ & $0.629_{-0.122}^{+0.105}$ & $0.326\pm0.076$ & -\tablefootmark{\small e} & multiple\\
CRISTAL-03 & $10.40\pm0.29$ & 4.8 & $0.044\pm0.014$ & $0.037\pm0.016$ & $10.86_{-0.25}^{+0.16}$ & $0.210_{-0.122}^{+0.077}$ & $\leq0.369$\tablefootmark{\small f} & $0.173\pm0.007$ & single\\
CRISTAL-04a & $10.15\pm0.29$ & 7.6 & $0.174\pm0.043$ & $0.164\pm0.035$ & $11.36_{-0.10}^{+0.08}$ & $0.587_{-0.162}^{+0.130}$ & $0.130\pm0.072$ & $0.154\pm0.007$ & multiple\\
CRISTAL-05 & $10.16\pm0.35$ & 6.7 & $0.141\pm0.044$ & $0.122\pm0.026$ & $11.36_{-0.11}^{+0.08}$ & $0.574_{-0.161}^{+0.129}$ & $0.244\pm0.105$ & $0.061\pm0.003$ & single\\
CRISTAL-06a & $10.09\pm0.30$ & 13.1 & $0.280\pm0.051$ & $0.254\pm0.029$ & $11.55_{-0.05}^{+0.05}$ & $0.794_{-0.122}^{+0.109}$ & $0.310\pm0.071$ & $0.214\pm0.010$ & multiple\\
CRISTAL-07ab & $10.00\pm0.33$ & 4.5 & $0.078\pm0.037$ & $0.085\pm0.019$ & $11.16_{-0.11}^{+0.09}$ & $0.331_{-0.089}^{+0.071}$ & $0.201\pm0.153$ & $0.146\pm0.008$ & multiple\\
CRISTAL-08 & $9.85\pm0.36$ & - & - & $\leq0.073$ & $\leq10.99$ & $\leq0.363$ & - & $0.380\pm0.012$ & single\\
CRISTAL-09 & $9.84\pm0.39$ & 4.5 & $0.083\pm0.032$ & $0.080\pm0.036$ & $11.18_{-0.26}^{+0.16}$ & $0.463_{-0.302}^{+0.189}$ & $0.104\pm0.079$ & $0.028\pm0.002$ & single\\
CRISTAL-10 & $9.99\pm0.31$ & 4.4 & $0.122\pm0.058$ & $0.101\pm0.026$ & $11.29_{-0.13}^{+0.1}$ & $0.333_{-0.104}^{+0.08}$ & $0.405\pm0.226$ & -\tablefootmark{\small e} & multiple\\
CRISTAL-11a & $9.68\pm0.33$ & 4.6 & $0.137\pm0.055$ & $0.095\pm0.051$ & $11.11_{-0.33}^{+0.19}$ & $0.542_{-0.475}^{+0.265}$ & $0.116\pm0.091$ & $0.066\pm0.004$ & single\\
CRISTAL-12 & $9.30\pm0.47$ & - & - & $\leq0.029$ & $\leq10.73$ & $\leq0.535$ & - & -\tablefootmark{\small g} & single\\
CRISTAL-13  & $9.65\pm0.34$ & 6.6 & $0.137\pm0.039$ & $0.120\pm0.037$ & $11.23_{-0.16}^{+0.12}$ & $0.621_{-0.268}^{+0.196}$ & $0.092\pm0.070$ & $0.056\pm0.023$ & multiple\\
CRISTAL-14 & $9.53\pm0.38$ & - & - & $\leq0.044$ & $\leq10.77$ & $\leq0.364$ & - & $0.054\pm0.011$ & single\\
CRISTAL-15 & $9.69\pm0.33$ & - & - & $\leq0.046$ & $\leq10.81$ & $\leq0.275$ & - & -\tablefootmark{\small e} & single\\
CRISTAL-16 & $9.60\pm0.39$ & - & - & $\leq0.037$ & $\leq10.85$ & $\leq0.386$ & - & $0.250\pm0.020$ & single\\
CRISTAL-17 & $9.51\pm0.40$ & - & - & $\leq0.035$ & $\leq10.83$ & $\leq0.362$ & - & -\tablefootmark{\small g} & single\\
CRISTAL-19 & $9.51\pm0.36$ & 5.5 & $0.093\pm0.038$ & $0.078\pm0.022$ & $11.13_{-0.15}^{+0.11}$ & $0.572_{-0.220}^{+0.165}$ & $0.262\pm0.153$ & $0.097\pm0.009$ & single\\
\hline
\multicolumn{9}{c}{{\bf CRISTAL+ sample}} \\
CRISTAL-20 & $10.11\pm0.35$ & 13.8 & $0.157\pm0.025$ & $0.144\pm0.015$ & $11.43_{-0.05}^{+0.04}$ & $0.459_{-0.055}^{+0.049}$ & $0.251\pm0.054$ & $0.060\pm0.002$ & single\\
CRISTAL-21 & $10.11\pm0.32$ & 4.9 & $0.528\pm0.219$ & $0.201\pm0.046$ & $11.54_{-0.11}^{+0.09}$ & $0.613_{-0.187}^{+0.148}$ & $1.160\pm0.499$ & $0.062\pm0.003$ & single\\
CRISTAL-22ab\tablefootmark{\small h} & $10.35\pm0.37$ & 22.9 & $1.522\pm0.074$ & $1.343\pm0.158$ & $12.42_{-0.05}^{+0.05}$ & $0.927_{-0.158}^{+0.140}$ & $0.150\pm0.011
$ & $0.129\pm0.008$ & multiple\\
CRISTAL-23a & $10.55\pm0.29$ & 12.2 & $1.142\pm0.211$ & $0.831\pm0.087$ & $12.07_{-0.05}^{+0.04}$ & $0.847_{-0.123}^{+0.111}$ & $0.486\pm0.093$ & $0.091\pm0.006$ & multiple\\
CRISTAL-24 & $10.53\pm0.08$ & 22.3 & $1.201\pm0.100$ & $1.236\pm0.092$ & $12.24_{-0.03}^{+0.03}$ & $0.956_{-0.102}^{+0.094}$ & $0.204\pm0.028$ & $0.105\pm0.020$ & single\\
CRISTAL-25 & $10.90\pm0.32$ & 7.8 & $0.501\pm0.137$ & $0.402\pm0.065$ & $11.75_{-0.08}^{+0.06}$ & $0.681_{-0.145}^{+0.123}$ & $0.320\pm0.116$ & $0.131\pm0.004$ & single\\
\hline
\multicolumn{9}{c}{{\bf serendipitously detected sources}} \\
CRISTAL-01b & $9.81\pm0.34$ & 4.4 & $0.449\pm0.165$ & $0.340\pm0.066$ & $11.67_{-0.09}^{+0.08}$ & $0.715_{-0.190}^{+0.156}$ & $0.331\pm0.153$ & $0.095\pm0.005$ & multiple\\
CRISTAL-07c & $10.21\pm0.35$ & 7.0 & $0.160\pm0.045$ & $0.123\pm0.024$ & $11.32_{-0.09}^{+0.08}$ & $0.683_{-0.181}^{+0.148}$ & $0.139\pm0.077$ & $0.143\pm0.015$ & single\\
\hline
\end{tabular}
\tablefoot{
\tablefoottext{a}{rest frame $158\mu{\rm m}$ fluxes measured with {\sc UVMultifit}.}
\tablefoottext{b}{rest frame $158\mu{\rm m}$ fluxes measured with tapered images (see Section \ref{subsec:dustsize}). 4-$\sigma$ upper limits are shown if source ${\rm S/N}<4$.}
\tablefoottext{c}{circularized {\it HST}/F160W size with fixing s{\`e}rsic index to unity.}
\tablefoottext{d}{classification based on the {\it HST}/F160W image.}
%\vspace{-6pt}\tablefoottext{\small d}{the stellar mass estimation may be misestimated due to the contamination of nearby SMG.}
\tablefoottext{e}{galfit does not converge due to the existence of multiple sources having similar brightness.}
\tablefoottext{f}{1-$\sigma$ upper limit of the size.}
\tablefoottext{g}{no clear counterpart or low S/N in {\it HST}/F160W image.}
\tablefoottext{h}{$S_{\rm uv}$ indicate total flux of main+sub components, and $r_{e, {\rm dust}}$/$r_{e, {\rm UV}}$ are the effective radius of the main component (See Figure \ref{fig:thumbneils}).}
}
\label{tab:tab1}
\end{table*}
\renewcommand{\arraystretch}{1}

%
%
%
% Figure 7
%
%
%
%
%
%
%
%
%

%
%
%
% Section 4
%
%
%
\section{Discussion}\label{sec:discussion}
\subsection{Fraction of obscured star formation}
The obscured fraction of star formation ($f_{\rm obs}$) is defined by the ratio of the dust-obscured star formation detected in rest-frame IR continuum emission to the total star formation rates accounting for dust-unobscured star formation observable in rest-frame UV continuum emission as following: 
\begin{equation}
f_{\rm obs}=\frac{{\rm SFR}_{\rm IR}}{{\rm SFR}_{\rm UV}+{\rm SFR}_{\rm IR}}
\end{equation}
In general, the obscured fractions show a positive correlation with the stellar masses of the galaxies \citep[e.g.,][]{2017ApJ...850..208W,2018ApJ...853...56R} as well as some studies presenting the same trend in IR excess (IRX; $\log_{10} L_{\rm IR}/L_{\rm UV}$), which is a similar quantity with $f_{\rm obs}$ as an indicator of the obscured star formation \citep[e.g.,][]{2009ApJ...698L.116P,2015ApJ...807..141P,2016ApJ...833...72B,2017MNRAS.466..861D,2018MNRAS.476.3991M,2020ApJ...902..112B}.
Based on the results in \citet{2017ApJ...850..208W}, the obscured fractions do not exhibit any clear evolution from $z\sim2.5$ to $z\sim0$ in the average obscured fraction of $f_{\rm obs}=0.8$ (for \citealt{2002ApJ...576..159D} template) or 0.6 (for \citealt{2012ApJ...760....6M} and \citealt{2015A&A...573A.113B} template).
Recent ALMA studies suggest a possible decrease in the mass dependence of $f_{\rm obs}$ at a high mass range ($M_{\ast}\gtrsim10^{10}\,M_{\odot}$) at $z>4$ \citep{2020MNRAS.491.4724F,2020A&A...643A...4F,2022MNRAS.515.3126I,2023MNRAS.518.6142A}. 

\begin{figure*}[htbp]
\centering
\includegraphics[width=18.5cm]{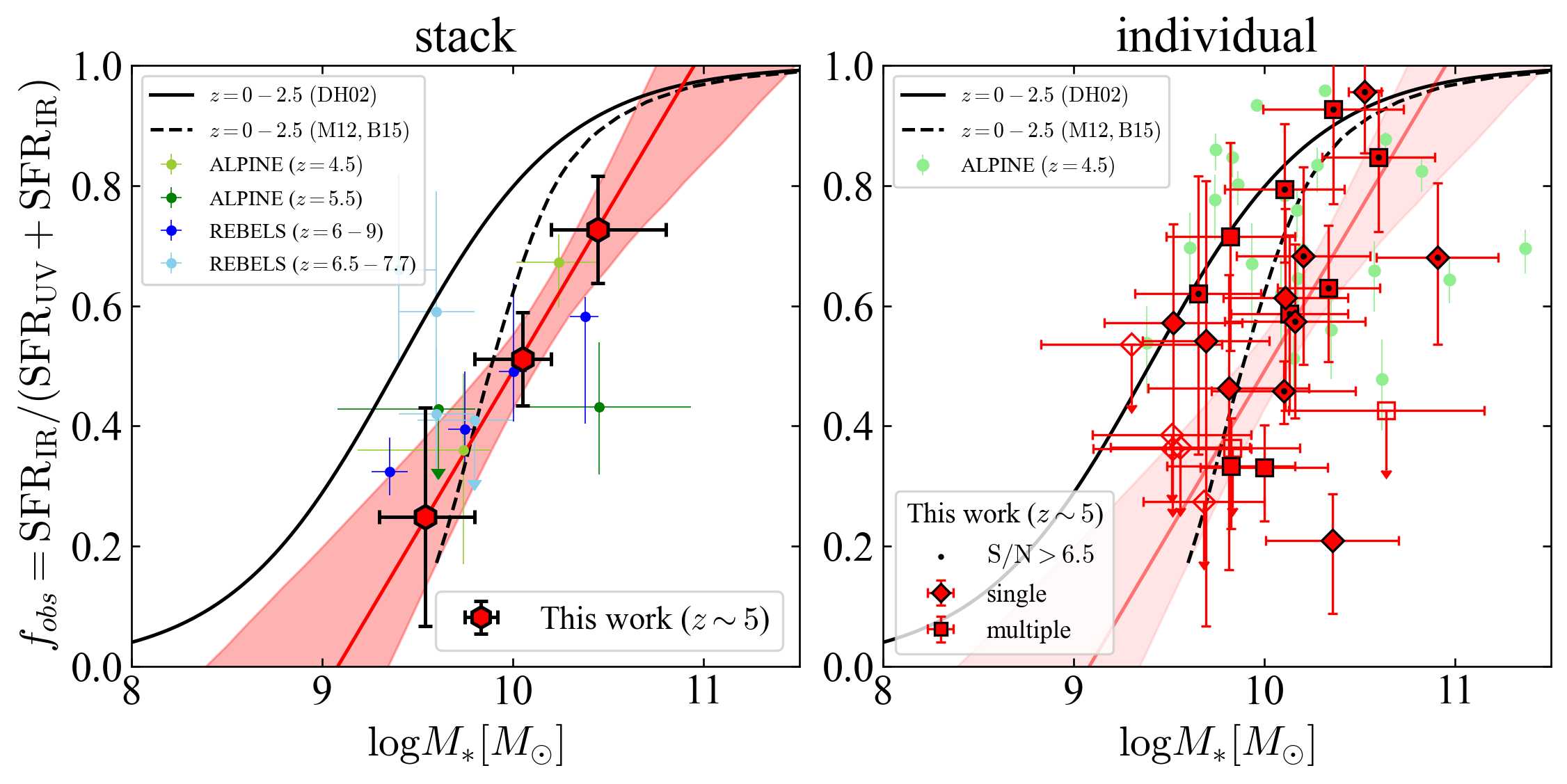}
\caption{\textbf{(left)} The relationship between the obscured fraction and the stellar mass from stacking analysis. Black solid and dashed lines show the results from \citet{2017ApJ...850..208W} with two different dust emission templates of \citet{2002ApJ...576..159D} template (D02) and \citet{2012ApJ...760....6M} and \citet{2015A&A...573A.113B} template (M12, B15). The obscured fraction increases as a function of the stellar mass, and the trend has a good agreement with the results at $z\sim5$ from the ALPINE survey \citep{2020A&A...643A...4F} and those at $z\sim6$--9 from the REBELS survey \citep{2023MNRAS.518.6142A}, while another result at $z\sim6.5$--7.7 from the REBELS survey \citep{2023arXiv230917386B} show an opposite trend with slightly higher $f_{\rm obs}$ at $M_{\ast}<10^{10}M_{\odot}$. \textbf{(right)} The obscured fraction of the individual galaxies as a function of the stellar masses. The diamonds and squares correspond to the two classifications of single and multiple. We also show the results of the galaxies detected in the dust continuum at $z\sim5$ \citep{2020A&A...643A...4F}. Our observations capture the galaxies with smaller $f_{\rm obs}$ compared with previous $z\sim5$ galaxies individually detected thanks to deep observations. There is a significant variation in $f_{\rm obs}$-$M_{\ast}$ relation from galaxy to galaxy.}
\label{fig:Mstarfobs}
\end{figure*}

In order to confirm if similar trends can be observed in our data, we calculate the average obscured fraction of the sample binned by their stellar masses.
We divide the sample into three ranges with $\log_{10} M_{\ast}\,[M_{\odot}]=[9.3: 9.8]$, [9.8: 10.2] and [10.2: 10.9] so that the number of galaxies in each bin is similar.
%We perform stacking analysis in the visibility domain by concatenating visibilities of the galaxies in each stellar mass bin with the {\sc CASA} task {\sc concat} and run {\sc UVMultifit} to measure total fluxes.
For the stacking analysis, we make the images with the taper scale of $2''$ after shifting the phase center of the image to the central position of the \cii\ emission (Ikeda et al. in prep.).
Then we apply the {\sc CASA}/{\sc imsmooth} to unify the beam size to $3''$ following the flux measurements in Section \ref{subsec:dustsize} and perform median stacking of the images.
We take into account the redshift difference of the target galaxies by multiplying the factor of $(1+z)/d_L^2$ to the image because our calculation from $S_{158\mu{\rm m}}$ to $L_{\rm IR}$ is proportional to the reciprocal of the factor.
Note that the offset between the dust continuum and \cii\ emission does not impact the result because the typical offset ($\sim0.2''$) is much smaller than $3''$ beam size.
%measured flux changes only \red{XX}\%.
We confirm that the results do not change significantly if we adopt the inverse $L_{\rm UV}$ weight following \citet{2020A&A...643A...4F}.
The infrared luminosity is derived from a peak flux of the dust continuum as in section \ref{sec:analysis}.
%Note that the stacking procedure used here is different from that in Section \ref{subsec:stack}, as the purpose of the stacking here is to obtain the average flux of each sample.
We calculate the median $L_{\rm UV}$ of the galaxies obtained from the best-fit SEDs in each bin.
Then we convert them into ${\rm SFR}_{\rm UV}$ or ${\rm SFR}_{\rm IR}$ following \citet{2014ARA&A..52..415M}. %\citet{1998ARA&A..36..189K,2012ARA&A..50..531K}.
Figure \ref{fig:Mstarfobs} (left) illustrates the average $f_{\rm obs}$ in each stellar mass bin ($f_{\rm obs}$-$M_{\ast}$).
Our results support that the obscured fraction at the range of $M_{\ast}<10^{10}\,M_{\odot}$ does not show clear evolution from $z=0$--2.5 (\citet{2012ApJ...760....6M} and \citet{2015A&A...573A.113B}) to $z\sim5$.
On the other hand, we find the possible evolution of the obscured fraction from $z\sim5$ to $z\sim0$--2.5 at the range of $M_{\ast}>10^{10}\,M_{\odot}$, and that implies a build-up of dust in massive galaxies.
Note that as our sample selection is originally based on a bright UV continuum and emission lines, it likely led to missing obscured dusty galaxies \citep[e.g.,][]{2018A&A...620A.152F,2019ApJ...878...73Y}.
We find our results have a very good agreement with the results at $z=4.5$ and $z=5.5$ from the ALPINE survey \citep{2020A&A...643A...4F}, and $z=6$--9 from the REBELS survey \citep{2023MNRAS.518.6142A}.
Compared with the another results at $z=6.5$--7.7 from the REBELS survey \citep{2023arXiv230917386B}, our results are indicate slightly lower $f_{\rm obs}$ at $M_{\ast}<10^{10}\,M_{\odot}$.
\citet{2023arXiv230917386B} discussed that the the effect of scatter in the obscuration or the clumpy morphology of their target galaxies may cause such a difference.
%and potentially show large $f_{\rm obs}$ at $M_{\ast}>10^{10}\,M_{\odot}$ .
%Given that the number density of galaxies with mass $M_{\ast}>10^{10}\,M_{\odot}$ increases from $z=6$--9 to $z=4$--6 \citep{2017A&A...605A..70D,2021ApJ...922...29S}, the cosmic dust-obscured star formation activity is also expected to increase from $z=6$--9 to $z=4$--6.
%Moreover, the consistent obscured fraction at $z=6$--9 and $z=4$--6 may indicate little evolution of the mass-metallicity relation ($M_{\ast}-Z$) during $z\sim4$--9 since interstellar dust originates from metal \citep[e.g.,][]{2007ApJ...663..866D,2013EP&S...65..213A}.
%Recent {\it JWST} observations also imply only a mild evolution of the $M_{\ast}-Z$ relation at $z>3$ \citep{2023arXiv230112825N,2023arXiv230408516C} while the stellar mass ranges are slightly lower than that covered in our study.

Figure \ref{fig:Mstarfobs} (right) shows the obscured fraction of the individual galaxies as a function of the stellar mass. 
The CRISTAL galaxies show a large diversity of $f_{\rm obs}=0.2$--0.9.
Since we utilized the same scaling factor as \citet{2020A&A...643A...4F} to convert $S_{158\mu{\rm m}}$ into $L_{\rm IR}$ \citep{2020A&A...643A...2B}, lower values of $f_{\rm obs}$ suggest smaller $S_{158\mu{\rm m}}$.
Some CRISTAL galaxies show smaller $f_{\rm obs}$ than galaxies individually detected in \citet{2020A&A...643A...4F}, with a median $f_{\rm obs}$ in \citet{2020A&A...643A...4F} of 0.70 compared to 0.61 in our work, respectively.
The smaller values of $f_{\rm obs}$ are consistent with those stacking analyses \citep{2020A&A...643A...4F,2023MNRAS.518.6142A}.
By utilizing the coverage of the wide range $f_{\rm obs}$, we discuss the variation of $f_{\rm obs}$ in the next section.

%
%
%
% Figure 8
%
%
%

\subsection{Origin of the spread in $f_{\rm obs}$-$M_{\ast}$ plane}
In Figure \ref{fig:Mstarfobs} (right), we found a large spread in the $f_{\rm obs}$-$M_{\ast}$ relation across galaxies in our sample. 
As more than 30\% of the galaxies deviate from the average relationship derived from the stacking analysis, even when considering 1$\sigma$ errors, the variance is not likely to come from the measurement errors.
Such a variation has been reported in ALMA observations toward UV-bright galaxies at $z>4$ \citep{2020A&A...643A...4F,2022MNRAS.515.3126I}.
They investigated the relationship between UV continuum slope ($\beta_{\rm UV}$) and IRX.
%While IRX-$\beta_{\rm UV}$ relation depends only on the dust attenuation law, there is a variation of IRX values in the same $\beta_{\rm UV}$.
Previous studies predict that the scatter could be caused by dust grain type \citep[e.g.,][]{2017MNRAS.471.5018F}, dust geometry \citep[e.g.,][]{2000ApJ...528..799W}, stellar populations
\citep[e.g.,][]{2017MNRAS.472.2315P}, stellar or dust assembly \citep[e.g.,][]{2017MNRAS.471.3152P}, spatial decoupling of UV and IR emission due to galactic interaction or dust-enshrouded nuclear starbursts \citep[e.g.,][]{2010ApJ...715..572H}, and inclination \citep[e.g.,][]{2018ApJ...869..161W}.
As it is difficult to demonstrate the dependence of dust/stellar properties such as grain type, geometry, or assembly based on our ALMA observations, we discuss the potential correlation between $f_{\rm obs}$ and morphology.
We note that no correlation is identified between $f_{\rm obs}$ and the stellar ages derived from SED fitting obtained in Section \ref{subsec:globsed}.

\subsubsection{inhomogeneous dust reddening}
The first possible explanation for the variation in the obscured fraction is inhomogeneous dust reddening, such as a global spatial offset between dust-enshrouded and dust-free star-forming regions.
Recently, \citet{2022MNRAS.510.5088B} discussed the impact of the spatial distribution of rest-UV and IR emission on the UV-IR energy balance at $z\sim7$ \citep[see also][]{2022MNRAS.515.3126I}.
They found a correlation between a redder UV slope and a peak of FIR emission detected in ALMA and suggest that inhomogeneous dust reddening might be responsible for variations in the spatially-integrated IRX-$\beta_{\rm UV}$ relation \citep[see also,][]{2015ApJ...813...36V,2018MNRAS.477..552B}.

In Figure \ref{fig:offsetfobs}, we show the distance from our best-fit $f_{\rm obs}$-$M_{\ast}$ relation ($\Delta f_{\rm obs}$) as a function of the spatial offset of the central position between the UV and dust continuum. 
We find a negative correlation between $\Delta f_{\rm obs}$ and the spatial offset within 1$\sigma$ uncertainty of the correlation coefficient of $\rho_{\rm corr}=-0.58_{-0.20}^{+0.30}$ with a $p$-value of 0.082. 
This correlation suggests that the decoupling of UV and IR emissions contributes to the escape of UV photons from interstellar space, and makes the variation on the obscured fraction.
%while higher accuracy of the central positions with deeper observations is necessary.

What causes the spatial offset between UV- and IR-bright star-forming regions?
Numerical simulations incorporating radiative transfer suggest that galactic interaction or feedback are possible mechanisms to decouple UV- and IR-bright regions \citep[e.g.,][]{2011MNRAS.412..411Y,2017ApJ...840...15S,2019MNRAS.488.2629A,2021MNRAS.502.3210L}.
%One is the galactic interactions.
%A detailed study of LIRGs indicates that the UV slope becomes decoupled from the IR emission during mergers \citep{2010ApJ...715..572H}.
For the feedback case, it is expected that galaxies with higher sSFR will tend to show larger peak offsets because star formation-driven outflows are more prevalent \citep[e.g.,][]{2020ApJ...900....1F,2020ApJS..247...61F}.
We divide the sample into two groups within ${\rm sSFR}<4\,{\rm Gyr}^{-1}$ and ${\rm sSFR}>4\,{\rm Gyr}^{-1}$ based on the total (IR+UV) SFR and compare the peak offset between UV and IR. 
There is no clear difference in the peak offset between the two groups, $0.150\pm0.046''$ for the galaxies within ${\rm sSFR}<4\,{\rm Gyr}^{-1}$ and $0.150\pm0.038''$ for ${\rm sSFR}>4\,{\rm Gyr}^{-1}$.

For the merger case, we expect a larger spatial offset in the ``multiple'' sample than the ``single'' sample because the galaxies classified as ``multiple'' correspond to a kind of early-stage merger.
%Since our galaxy classifications are based only on the 2D appearance in {\it HST} images,
%On the other hand, it is difficult to distinguish whether the ``single'' galaxy is an isolated disk galaxy or a late-stage merger.
However, the peak offset between UV and IR does not exhibit a systematic offset between ``multiple'' ($0.145\pm0.048''$) and ``single'' ($0.155\pm0.041''$).
Therefore the interaction is not likely to primary cause of the decoupling.
%Numerical simulations incorporating radiative transfer suggest that feedback by supernovae or AGN, which could make UV and IR emissions decouple, contribute to the escape of UV photons from interstellar space \citep[e.g.,][]{2011MNRAS.412..411Y,2014MNRAS.440..776Y,2019MNRAS.488.2629A}.
%The physical origin of the decoupling of UV and IR is unclear in our dataset.
While the merger activity does not generate the decoupling of the dust-obscured and unobscured star-forming regions, it may accelerate the obscured star formation.
In Figure \ref{fig:Mstarfobs}, we plot our result separately for single (isolated) and multiple (merging) galaxies.
Most highly obscured galaxies tend to have high total SFRs and show merging signatures in {\it HST} images. 
Also, The average obscured fraction and SFR of the multiple galaxy sample ($\langle f_{\rm obs} \rangle_{\rm multiple} = 0.43^{+0.03}_{-0.03}$, $\langle\, \log_{10} {\rm SFR}\,[M_{\odot}\,{\rm yr}^{-1}] \rangle_{\rm multiple} = 1.97^{+0.60}_{-0.54}$) is higher than that of the single galaxy samples ($\langle f_{\rm obs} \rangle_{\rm single} = 0.34^{+0.03}_{-0.03}$, $\langle\, \log_{10} {\rm SFR}\,[M_{\odot}\,{\rm yr}^{-1}] \rangle_{\rm single} = 1.58^{+0.10}_{-0.11}$).
Our results suggest that galaxy interactions may boost not only the star formation activities but also the obscured fraction within almost the same stellar mass range, and the mixture of the different merger stages and isolated disks leads to the variety in $f_{\rm obs}$-$M_{\ast}$ plane.
%Note that since our galaxy classifications are based only on the 2D appearance in {\it HST} images, the classifications may not be entirely accurate.
%The second possible explanation for the variation in the obscured fraction is a global spatial offset between dust-enshrouded and dust-free star-forming regions.

Another potential physical reason for the displacement is a multiphase structure of the ISM, for which one-zone ISM models assumed in the IRX-$\beta$ (or $f_{\rm obs}$-$M_{\ast}$) relation cannot be applied. 
As discussed in \citet{2022MNRAS.512...58F}, {\it molecular index} $I_{\rm m}$ ($=F_{158}/F_{1500}/(\beta-\beta_{\rm int})$) might represent a good indicator of such a multi-phase ISM structure. 
We test the hypothesis by comparing the UV-dust spatial offset and $I_{\rm m}$ under the assumption of $\beta_{\rm int}=-2.41$ following \citet{2022MNRAS.515.3126I}, but we do not find a clear correlation between them ($\rho_{\rm corr}=-0.218_{-0.234}^{+0.262}$ with a $p$-value of 0.401). This result is mainly due to the small measured values of the UV-dust offset compared to \citet{2022MNRAS.515.3126I}. 
Interestingly, though, CRISTAL-24 shows a very high $I_{\rm m}$ of $\sim5000$ with a relatively blue UV slope $\beta=-1.55$. 
An inconsistent ${\rm SFR}_{\rm SED}$ and ${\rm SFR}_{\rm UV+IR}$ in CRISTAL-24 described in Section \ref{subsec:globsed} may indicate multi-phase and component nature of the CRISTAL-24.

\subsubsection{concentration of the star-forming regions}
Since we do not find clear dependence on the potential interactions or UV-dust offsets, we will now compare $f_{\rm obs}$ with the compactness of the star-forming regions.
\citet{2017ApJ...850..208W} indicate that the spatial extent of the star-forming region affects the obscured fraction \citep[see also,][]{2021MNRAS.502.3426S}.
\citet{2010ApJ...715..572H} mention that dust-enshrouded central starburst may lead to a diversity of the obscured fraction.
%As the compactness could be affected by the contamination of nearby companions to the size measurements in the ``multiple'' galaxies, here we focus on the ``single'' galaxies with ${\rm S/N}\geq6.5$ in the dust continuum to obtain robust results.
We derive the compactness of the dust-obscured star-forming region as $\Sigma_{\rm IR}$ computed from infrared luminosities and effective radii of the dust continuum.
We also check the dependence of $\Delta f_{\rm obs}$ on the size ratio between UV and dust emission ($r_{e, {\rm dust}}/r_{e, {\rm UV}}$).
We find that there is no clear correlation between $\Delta f_{\rm obs}$ and the size ratio ($\rho_{\rm corr}-0.11_{-0.27}^{+0.28}$ with a $p$-value of 0.693 respectively) but $\Sigma_{\rm IR}$ has potential correlation with $f_{\rm obs}$ ($\rho_{\rm corr}={0.30}_{-0.26}^{+0.22}$ with a $p$-value of 0.245).
Enhancement of the dust-obscured star formation activity could relate to the concentration of the young, active dust-enshrouded star formation regions, which may be accelerated by unstable disks with high gas mass fraction \citep{2009ApJ...703..785D,2013ApJ...768...74T,2017ApJ...837..150S,2020A&A...643A...5D}.
%Furthermore, we find the same trend if we include ``multiple'' galaxies. 
Since starbursts induced by the galaxy interaction could concentrate the gas and dust into the central parts of galaxies \citep[e.g.,][]{1996ApJ...471..115B}, it may accelerate the dust obscuration of the galaxy and could explain the positive trend between the obscured fraction and $\Sigma_{\rm IR}$ or merging signatures.

\subsubsection{inclination}
Both observations \citep[e.g.,][]{2018A&A...616A.157L} and simulations \citep[e.g.,][]{2010MNRAS.403...17J} studies suggest that an apparent UV-IR energy balance can be affected by a viewing angle of disk galaxies i.e., inclination.
There is a systematic offset of $\sim0.5$ dex in IRX values between edge-on and face-on galaxies \citep{2018ApJ...869..161W}.
While the inclinations are calculated from the axis ratios, the axis ratios of the UV continua may not be a proper indicator of an inclination due to severe dust attenuation.
We do not find a correlation between the axis ratio of the UV continuum and $\Delta f_{\rm obs}$ ($\rho_{\rm corr}=-0.09_{-0.27}^{+0.28}$ with a $p$-value of 0.761).
Moreover, only a few CRISTAL galaxies have constraints on the axis ratios in dust emission. 
The current data quality is not sufficient to discuss the possibility that inclinations explain the variation in $f_{\rm obs}$.
It is worth noting that two galaxies with relatively large $f_{\rm obs}$ ($>0.7$) have axis ratios of $\sim0.2-0.3$ in the dust emission, suggesting that an edge-on view of disk galaxies may make the obscured fraction higher.
Also, note that the axis ratio of {\sc [Cii]} emission also possibly correlates with $\Delta f_{\rm obs}$ with $\rho_{\rm corr}=-0.35_{-0.22}^{+0.27}$ and a $p$-value of 0.203, where the axis ratio of {\sc [Cii]} is measured with the similar way as this work (Ikeda et al. in prep).
Future {\it JWST} observations will unveil rest-frame optical light distribution, which traces stellar mass distributions. 
It will allow us to investigate the inclination dependence of $f_{\rm obs}$ toward high redshift galaxies \citep[e.g.,][]{2023arXiv231008829C}.

%
%
%
% Figure 9
%
%
%
\begin{figure*}[htbp]
\centering
\includegraphics[width=18cm]{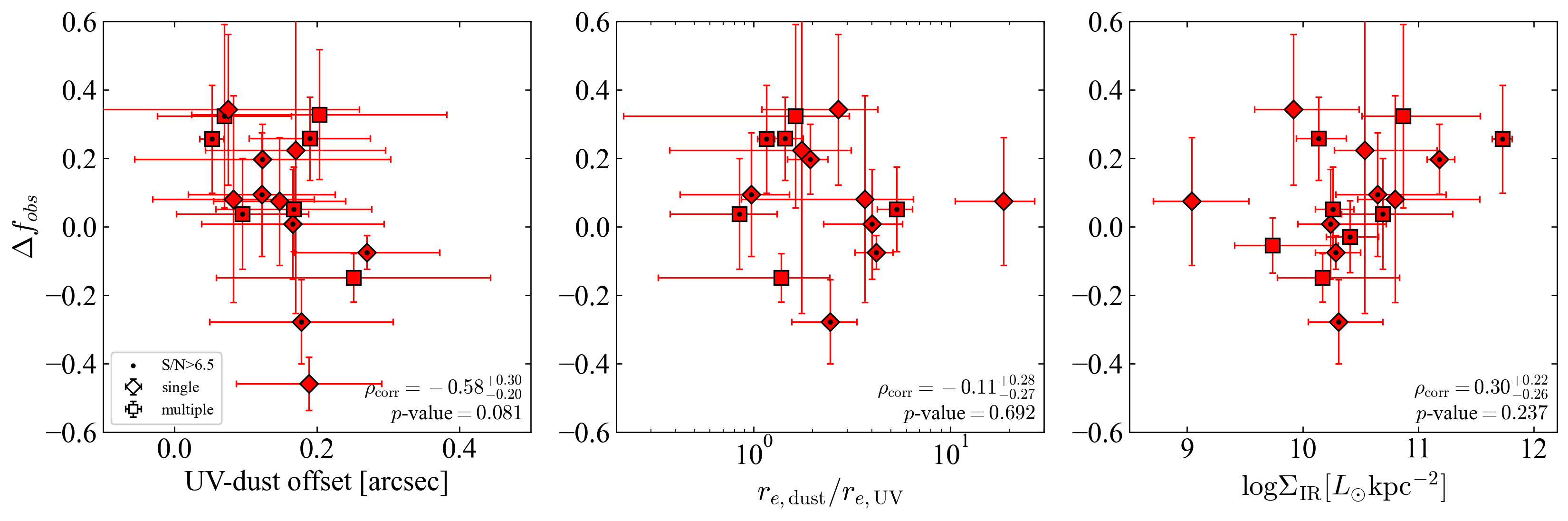}
\caption{The distance from an average $f_{\rm obs}$-$M_{\ast}$ relation as a function of (left) the spatial offset of the central position between UV and dust emission, (middle) the size ratio between UV and dust emission, and (right) the surface density of $L_{\rm IR}$. 
Galaxies without clear main UV components and without dust continuum detection at $<4\sigma$ are not shown. 
Based on the Spearman correlation coefficients and $p$-values, $f_{\rm obs}$ has a correlation with the UV-dust spatial offset and $\Sigma_{\rm IR}$.
On the other hand, there is no correlation between $f_{\rm obs}$ and the size ratio between UV and dust emission.}
\label{fig:offsetfobs}
\end{figure*}

\subsubsection{Caveat}
One caveat of our analysis is that we use the single template of FIR SEDs for all CRISTAL galaxies. 
Therefore we do not take into account variation of $T_{\rm dust}$ or $\beta_{\rm dust}$ characterizing FIR SEDs.
If actual $T_{\rm dust}$ of the galaxies having low (high) $f_{\rm obs}$ are larger (smaller) than that of the FIR SED used in this work \citep[$T_{\rm dust}\sim40$K, see][]{2020A&A...643A...2B}, the spread of $f_{\rm obs}$ becomes smaller.
A variation of $T_d\sim20\,{\rm K}$ changes infrared luminosities with $\Delta L_{\rm IR}\sim0.25$ dex and the obscured fractions with $\Delta f_{\rm obs}\sim0.15$.
Since we find a potential correlation between $\Sigma_{\rm IR}$ and $f_{\rm obs}$, the galaxies with high (low) $f_{\rm obs}$ can have higher (lower) $T_{\rm dust}$ or $L_{\rm IR}$ than the current estimation if we assume there is a positive correlation between $\Sigma_{\rm IR}$ and $T_{\rm dust}$ revealed in previous results \citep[e.g.,][]{2017ApJ...846...32D,2020MNRAS.494.3828D}.
In this case, the range of variation expands beyond the current estimation.
%This implies the variation cannot be fully explained by the variation of FIR SEDs.
We need high-frequency ALMA bands (Band-8/9/10), which cover the frequencies corresponding to the peak of the dust continuum emission, to derive the FIR SEDs of individual galaxies.
While high-frequency band observations towards large galaxy samples are still challenging due to limitations such as sensitivities or atmospheric conditions, case studies of some representative galaxies will help us to know accurate FIR SED.

\subsection{Star formation mode of $z\sim5$ LBGs}
\subsubsection{Extended dust-obscured star formation}\label{subsubsec:exdust}
%As described in Section \ref{subsec:UVsize} and Figure \ref{fig:reuvreir}, a comparison of effective radii between dust and UV emission for CRISTAL galaxies.
%In Figure \ref{fig:reuvreir}, we show a comparison between $r_{e, {\rm dust}}$ and $r_{e, {\rm UV}}$ for 13 CRISTAL galaxies. 
As described in Section \ref{subsec:UVsize} and Figure \ref{fig:reuvreir}, we find $r_{e, {\rm dust}}$ are comparable or slightly larger than $r_{e, {\rm UV}}$.
%whereas dust emission originates from the dust heated by UV radiation from massive O/B-type stars.
Our size ratios are not consistent with the results from TNG50 simulation expecting $r_{e, {\rm UV}}/r_{e, {\rm dust}}\sim2$--4 in the stellar mass range of the CRISTAL galaxies \citep{2022MNRAS.510.3321P}.
As \citet{2022MNRAS.510.3321P} measure half-light radii of UV and dust continuum at the almost same observed wavelength with our observations ($\lambda_{\rm obs}=1.6\mu{\rm m}$ and $850\mu{\rm m}$ for UV and dust continuum, respectively), the size ratios can be directly compared.

Before comparing observational results and the simulation, we need to be careful about observational biases coming from the limited S/Ns or spatial resolutions.
%As mentioned in Section \ref{subsubsec:MCsim}, the sizes of the CRISTAL galaxies whose dust continua are detected in $4.4\leq{\rm S/N}\leq5.5$ may be overestimated.
One possible reason for the extended dust continua is the source blending of nearby companions.
To demonstrate how the companions affect the size measurement, we re-run {\sc galfit} to the {\it HST} images after matching {\it HST} PSF to ALMA beam size with {\sc photutils/psf} and adding white noises to match peak S/Ns of UV continuum to dust continuum. 
As ALMA beam sizes can be changed by different weighting of visibilities, here we use natural-weighted beam sizes of each visibilities as nominal target beam sizes of the PSF matching.
Typically the naturally-weighted ALMA beam sizes ($0.46''=2.9\,{\rm kpc}$ at $z=5$) are larger than {\it HST} PSFs ($0.20''=1.3\,{\rm kpc}$ at $z=5$).
%, these PSF matching corresponds to the smoothing procedure.
We test the procedure in three CRISTAL galaxies (CRISTAL-02, 06a, and 22ab) since (i) they have nearby companions in {\it HST} images and (ii) their dust continua are detected in ${\rm S/N}>6.5$. 
The results reveal the measured sizes are consistent with those without PSF and S/N matching. 
Therefore source blending is not enough to explain the extended dust continua in CRISTAL galaxies.
Another possible reason is the limited S/N, but it is not likely to be a possible reason because $r_{e, {\rm dust}}$ are still only slightly larger than $r_{e, {\rm UV}}$ when we only take the sources with ${\rm S/N}>6.5$ into account.
%We note that because we use the visibility-based measurement for the dust continuum, the UV continuum sizes measured in smoothed images are not necessarily suitable for direct comparison.

%
%
%
% Figure 10
%
%
%
\begin{figure}[htbp]
\centering
\includegraphics[width=9cm]{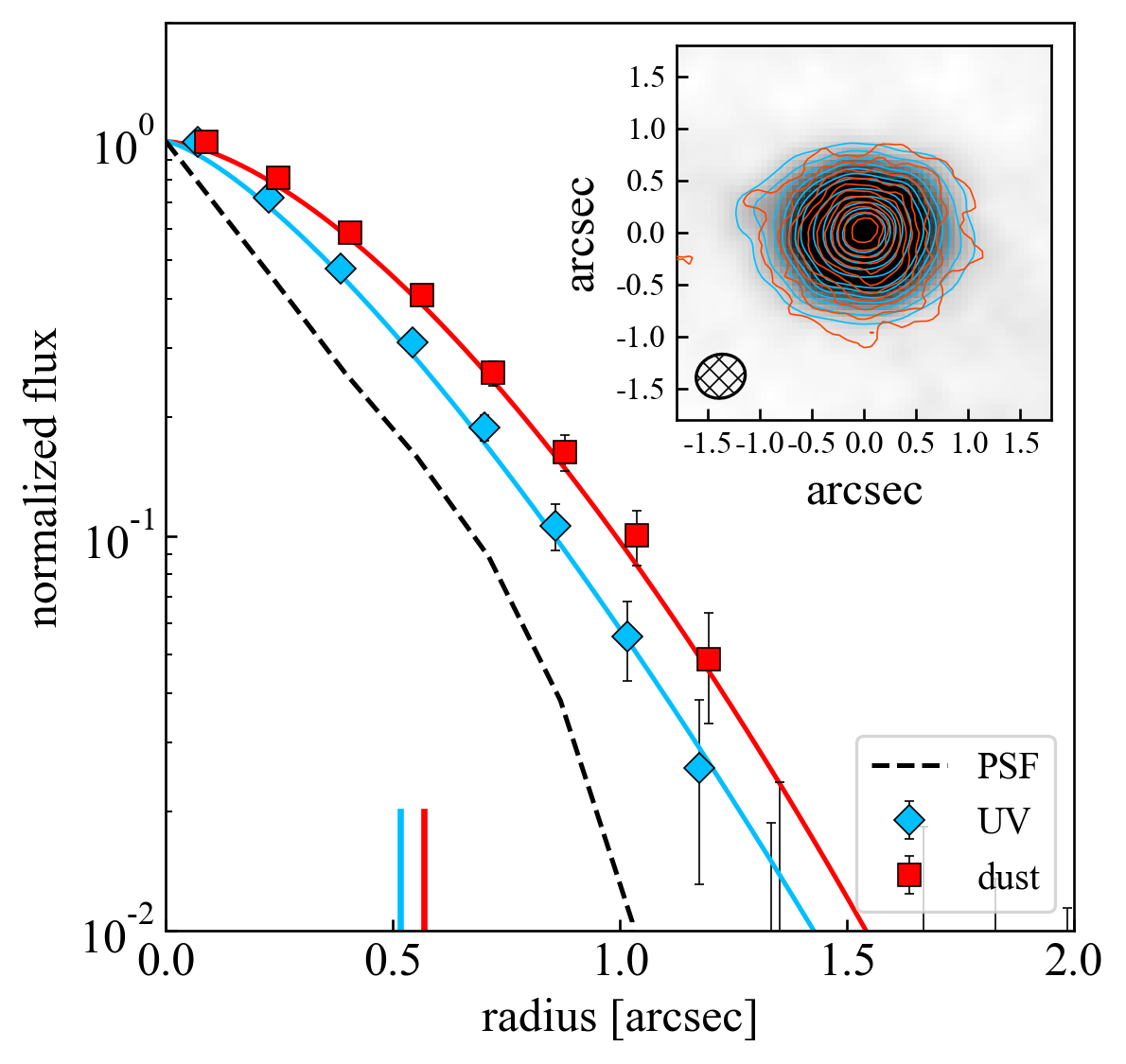}
%\captionsetup{labelformat=empty,labelsep=none}
\caption{Normalized radial profiles of the stacked dust and UV continua (red and blue respectively). The perpendicular lines are effective radii derived from the best-fit S{\'e}rsic profile in the corresponding colors. Here we stack only single (``isolated'') galaxy samples to avoid the blending of the companion. The matched PSF size is also shown in a black solid line. The inset shows the 2D distributions.  The spatial extent of the dust continuum is more extended to that of the UV continuum.}
\label{fig:radprof}
\end{figure}

In Figure \ref{fig:radprof}, we show the stacked radial profile of the dust and UV continuum to obtain better S/N than the individual measurements.
Here we stacked UV and dust continuums of single (``isolated'') galaxy samples to avoid blending with companions.
We follow the visibility-based method described in Section \ref{subsec:stack} regarding the stacking of ALMA data.
About the stacked UV continuum image, we shift the fitting center to the central coordinate of the image and match the PSFs of each {\it HST} image to that of the stacked ALMA image, and then take an average.
%Then we take an average with an inverse weight of the total flux estimated by {\sc galfit}.
Again, the stacked radial profile shows that the spatial extent of the dust continuum is slightly more extended than that of the UV continuum.

Therefore, the spatially extended dust continuum does not originate from the observation, but from the properties of the galaxy itself.
Possible physical interpretations of $r_{e, {\rm UV}}/r_{e, {\rm dust}}\lesssim1$ are clumpy gas distributions that enable UV photons to radiate molecular clouds out to the outskirts of the galaxies, even though these clouds are slightly distant from young massive stars.
The theoretical and observational studies suggest an increase of clumpy star-forming galaxies at high redshift \citep[e.g.,][]{2008ApJ...687...59G,2009ApJ...703..785D,2014ApJ...780...77T,2015ApJ...800...39G}.
In this case, the dust temperature is expected to decrease from the center to the outskirts of the galaxy.
%the total infrared luminosity decreases at the lower $T_{\rm dust}$ at the fixed continuum flux at rest-frame 158 $\mu$m, and the actual spatial extent of the dust-obscured star formation becomes compact.
In our conversion from $S_{\rm 158\mu{\rm m}}$ to $L_{\rm IR}$, if the dust temperature has a negative gradient of $\Delta T_{\rm dust}\sim10\,{\rm K}$ from center to the outskirts \citep[e.g.,][]{2019MNRAS.488.2629A}, the infrared luminosity decrease $\sim0.3\,{\rm dex}$, and the actual spatial extent of the dust-obscured star-forming region becomes comparable with that of the dust-unobscured ones as predicted in SERRA simulation \citep[e.g.,][]{2022MNRAS.513.5621P}.
Another possibility is that rest-UV observations do not trace global star formation activities but massive star-forming clumps within the galaxies.

\subsubsection{Stellar mass-size relation}
Several {\it HST}-based studies of galaxy morphology reveal that the size of the stellar distribution depends on both redshift and stellar mass \citep[e.g.,][]{2014ApJ...788...28V,2016ApJ...821...72S}. 
The stellar sizes become smaller with increasing redshift at fixed stellar mass and with decreasing stellar mass at fixed redshift. 
\citet{2017ApJ...850...83F} suggest dust continuum sizes also follow a similar evolutionary trend with redshift and stellar mass. 
Because the dust continuum traces recent star formation events and the stellar emission traces histories of the star formation in the galaxies, the size comparison of the dust continuum and stellar continuum at similar epochs brings a crucial perspective on the morphological transformation of galaxies.  
%These massive galaxies are classical SMGs or DSFGs because star formation activities of massive galaxies with $M_{\ast}\sim10^{11}\,M_{\odot}$ at $z\sim2$ are mostly dust-obscured \citep{2017ApJ...850..208W}.

\begin{figure}[htbp]
\centering
\includegraphics[width=9cm]{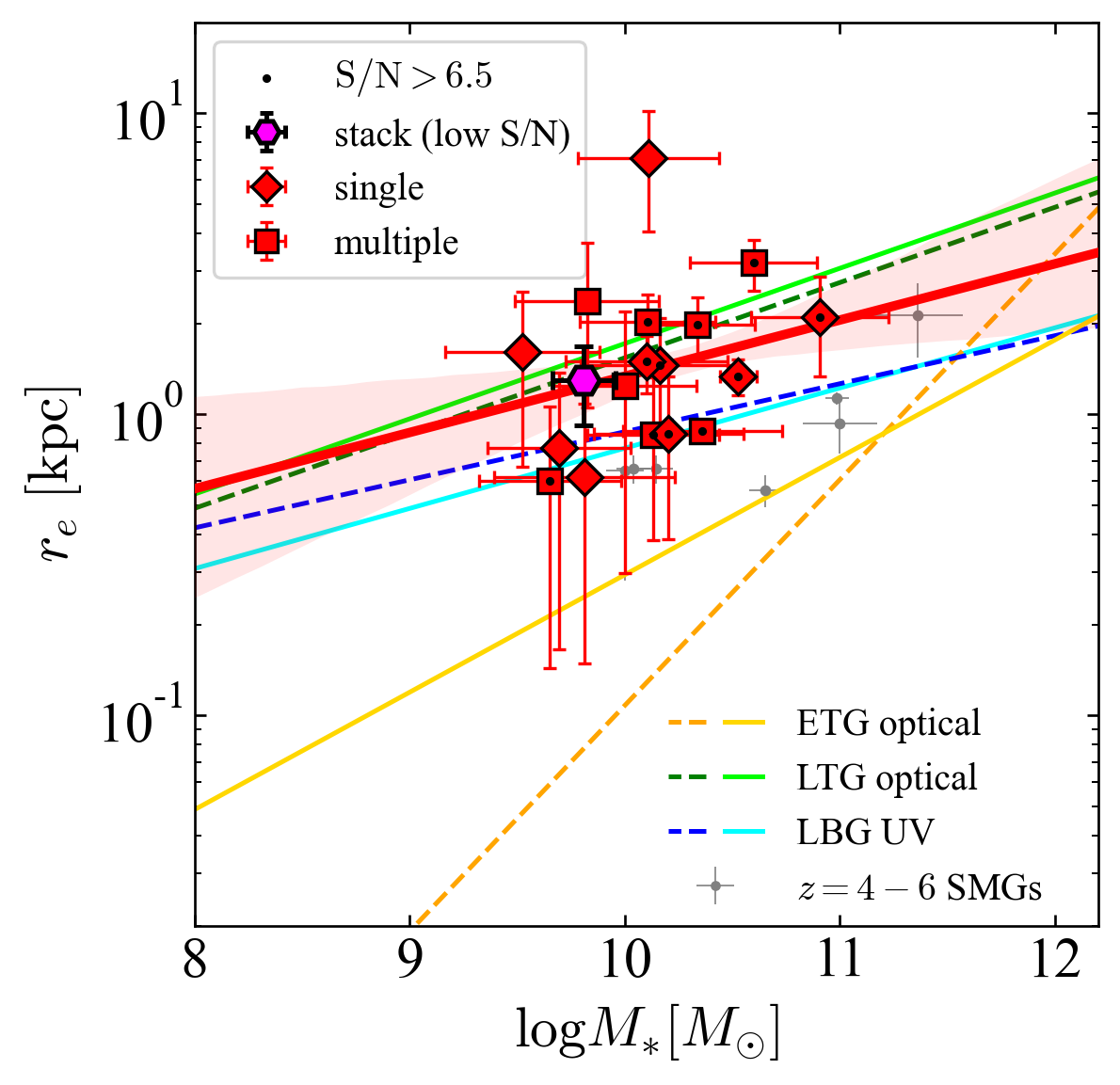}
\caption{Stellar mass versus effective radii of the dust continuum for the CRISTAL galaxies. The dashed and solid lines show the scaling relations derived with {\it HST} and {\it JWST}. The orange and green colors represent the rest-frame optical sizes of ETGs \citep{2014ApJ...788...28V,2023arXiv230706994I}, and the LTGs \citep{2014ApJ...788...28V,2023arXiv231102162W}. The blue colors indicate rest-frame UV sizes of LBGs \citep{2016ApJ...821...72S,2023arXiv230805018M}. The average stellar mass and stacked size derived in Section \ref{subsec:stack} is also shown. CRISTAL galaxies are distributed around the scaling sequence of LTGs or LBGs in contrast to the massive SMGs that show consistent dust continuum sizes and stellar masses with those of ETGs \citep{2014ApJ...796...84R,2014A&A...565A..59D,2015ApJ...798L..18H,2018ApJ...856..121G,2019ApJ...887...55C,2020ApJ...889..141T}.}
\label{fig:masssize}
\end{figure}

In Figure \ref{fig:masssize}, we plot rest-frame $158\mu{\rm m}$ dust continuum sizes of the CRISTAL galaxies as a function of their stellar masses. 
We also show typical rest-frame optical sizes of late-type galaxies (LTGs) and early-type galaxies (ETGs) \citep{2014ApJ...788...28V} and rest-frame UV sizes \citep{2016ApJ...821...72S} of Lyman-break galaxies (LBGs) as a function of stellar masses at $z\sim5$. 
Note that we extrapolated the redshift evolution reported in \citet{2014ApJ...788...28V} to $z\sim5$ because the rest-frame wavelength of 5000\,\AA\ used to measure stellar sizes in \citet{2014ApJ...788...28V} redshifted to the outside of the coverage of {\it HST} NIR filters. 
Therefore we also show recent {\it JWST} studies of the rest-frame UV sizes of LBGs \citep[][see also \citealt{2023arXiv230902790O}]{2023arXiv230805018M}, and those of the rest-frame optical sizes of LTGs \citep{2023arXiv231102162W} and ETGs \citep{2023arXiv230706994I}.
%and reveal the rest-frame UV and optical sizes are comparable at $z\sim4$--10 \citep{2023arXiv230902790O}.
The extrapolated \citet{2014ApJ...788...28V} relation might be slightly overestimated, although the difference does not impact our conclusion.
Future direct comparison with the rest-frame optical sizes measured by {\it JWST} observations will allow us to capture the dust, UV, and stellar size properties of the typical galaxies individually.

The dust continuum sizes and stellar masses of more dusty galaxies follow the trend of ETGs \citep[e.g.,][]{2022A&A...658A..43G}. 
These results imply that massive dusty galaxies are in a build-up phase of central stellar cores (i.e., stellar bulge) of massive elliptical galaxies \citep[e.g.,][]{2014ApJ...782...68T,2022A&A...658A..43G} or evolve into compact quiescent galaxies \citep[e.g.,][]{2014ApJ...782...68T,2015ApJ...810..133I,2016ApJ...827L..32B}.
This is consistent with the studies that investigated the stellar and dust continuum sizes of individual galaxies at $z\sim2$--3 \citep{2015ApJ...799...81S, 2019MNRAS.490.4956G,2019ApJ...879...54L,2020ApJ...901...74T}.
%At $z\sim2$, \citet{2019ApJ...879...54L} investigate stellar mass distributions of the classical SMGs, which are $S_{870\mu {\rm m}}\gtrsim 4$\,mJy, considering dust extinction effect. 
%They suggest that classical SMGs have centrally concentrated dust-obscured star formation activities within their extended stellar disks \citep[see also][]{2015ApJ...799...81S, 2019MNRAS.490.4956G}.
%\citet{2020ApJ...901...74T} measure dust and stellar spatial extent of uniformly selected massive DSFGs with $M_{\ast}\geq10^{11}\,M_{\odot}$ at $z\sim2$ and find compact dust-obscured star-forming regions in comparison to the extent of their stellar disks. 
%\citet{2016ApJ...827L..32B} also suggest strong nuclear starbursts, leading to the transformation of large star-forming galaxies into compact galaxies occur in massive galaxies with $M_{\ast}\sim10^{11}\,M_{\odot}$ at $z=2$--3 \citep[see also, ][]{2015ApJ...807..128S,2020ApJ...901...74T}. 
%\citet{2015ApJ...807..128S} argue that the compact nature of dusty star formation in SMGs corresponds to the bulge-growing phase of the compact QGs \citep[see also, ][]{2020ApJ...901...74T}.
In contrast, as shown in Figure \ref{fig:masssize}, the dust continuum sizes of CRISTAL galaxies follow a similar trend as the mass-size relation of LTGs or LBGs. 
Average stellar mass ($M_{\ast}\sim10^{9.8}\,M_{\odot}$) and effective radius ($\sim$1.7\,kpc) are in excellent agreement with them on the mass-size plane.
Our result suggests that typical star-forming galaxies with $M_{\ast}\sim10^{10}\,M_{\odot}$ might be still growing their disks via global star formation.

%
%
%
% Figure 9
%
%
%
%
%
%
%
%
%

%\section{MS galaxies as the pre-starburst phase}
\subsection{Morphological transformation and descendant}
Recent wide field surveys revealed typical dark matter halo masses of the bright LBGs ($M_{\rm UV}\sim-21\,{\rm mag}$, $M_{\ast}\sim10^{10}\,M_{\odot}$ in stellar mass) at $z=4$--6 are $M_{\rm halo}\sim10^{12}\,M_{\odot}$ \citep[e.g.,][]{2018PASJ...70S..11H}. 
If we assume the median growth rate for their dark matter haloes from N-body simulations, dark matter haloes with masses of $M_{\rm halo}\sim10^{12}\,M_{\odot}$ at $z\sim5$ evolve into haloes with $M_{\rm halo}\sim10^{12.5-13.0}\,M_{\odot}$ at $z\sim2$, which host massive galaxies with mass of $M_{\ast}\sim10^{11}\,M_{\odot}$ at $z\sim2$. 
In the case that the bright LBGs at $z=4$--6 continue to stay MS to $z=2$, their stellar masses could be $M_{\ast}\gtrsim10^{11}\,M_{\odot}$.
From these estimation, the bright LBGs with $M_{\ast}\sim10^{10}\,M_{\odot}$ at $z=4$--6 are likely to evolve into $z\sim2$ massive galaxies with $M_{\ast}\sim10^{11}\,M_{\odot}$.

As the average structural evolutions of SFGs are expected to follow the scaling relation on the mass-size plane, a bright LBG at $z=4$-6 will grow its stellar disk by a factor of two after a ten-fold increase in stellar mass.
%$\sim2\times$ larger effective radii during stellar masses of the bright LBGs at $z=4$--6 increase $\sim10\times$. 
The expected size of the stellar disks is consistent with that of massive DSFGs at $z\sim2$.
%The expansion of the stellar disks and following compaction events provide a good explanation for the conditions observed in massive DSFGs \citep{2016ApJ...827L..32B,2019ApJ...879...54L,2019ApJ...883...81S,2020ApJ...901...74T}. 
This consistency also supports that the bright LBGs at $z=4$--6 are expanding their stellar disks through widespread star formation.

\section{Summary and Conclusion}\label{sec:summary}
In this paper, we have focused on the dust continuum emissions of $z\sim5$ normal star-forming galaxies as a part of the ALMA cycle-8 large program, CRISTAL.
We have examined the spatial extent of the dust-obscured and unobscured star formation, obscured fraction of star formation as a function of stellar mass ($f_{\rm obs}-M_{\ast}$) relation, and origin of the observed variation in fobs with stellar mass on $f_{\rm obs}-M_{\ast}$.
The sample is representative of typical star-forming galaxies at $z=4$--6 as almost all of them lie within 0.3 dex of the main sequence of star-forming galaxies in that redshift range. 
The stellar mass ($M_{\ast}$) and SFR of the target galaxies are the range of $\log_{10} M_{\ast}\,[M_{\odot}]\sim9.5$--11.0 and $\log_{10} {\rm SFR}\,[M_{\odot}\,{\rm yr}^{-1}]\sim1.0$--2.7. 
From individual measurements and stacking analysis, we found:
\begin{enumerate}
\item The CRISTAL galaxies have $\sim$1 dex smaller infrared luminosities ($L_{\rm IR}$) and are slightly more extended in dust continuum emission than DSFGs or SMGs. Consequently the median IR surface densities $\Sigma_{\rm IR}$ are $\sim$10 times smaller (Figure \ref{fig:LvsR}). 
This suggests that representative galaxy population at $z\sim$4--6 undergoes moderate star formation with a lower $\Sigma_{\rm IR}$ compared with the dusty galaxies indicating intensive, centrally concentrated star formation.

\item The obscured fraction has a positive correlation with the stellar mass (Figure \ref{fig:Mstarfobs}, left).
Our averaged relationship is consistent with the previous results at $z\sim$5--9 and supports that the obscured fraction at the range of $M_{\ast}<10^{10}\,M_{\odot}$ does not show clear evolution from $z=0$--2.5, but may decrease at the range of $M_{\ast}>10^{10}\,M_{\odot}$.
The possible evolution at the range of $M_{\ast}>10^{10}\,M_{\odot}$ suggests the evolution of dust content from $z\sim5$ to $z\sim0$--2.5.

\item A large variation in $f_{\rm obs}$-$M_{\ast}$ relation across galaxies in our sample.
Through deep observations in our survey, we trace the galaxy with an obscured fraction of as low as 20\%, which is smaller than what has been reported in previous surveys.
We find a weak correlation between $f_{\rm obs}$ and potential merger identified by multiplicity, and compactness of the dust-obscured star-forming regions, but do not detect a correlation with the spatial offset between UV and dust continuum, or viewing angle (Figure \ref{fig:offsetfobs}).
Some mechanisms such as interactions or rich gas accretions may enhance the concentration of gas and dust, and accelerate the obscuration of the star formation activities.

\item The effective radii of the dust continuum ($r_{e, {\rm dust}}$) are comparable or $\sim2$ times larger than that of the UV continuum ($r_{e, {\rm UV}}$) in the individual comparison whereas the dust emission originates the dust heated by UV radiation from massive O/B type stars.
These results are not consistent with the Illustris TNG50 simulation which expects $r_{e, {\rm dust}}/r_{e, {\rm UV}}\sim0.25$--0.5 in the stellar mass range of the CRISTAL galaxies.
With careful treatment and stacking analysis, we confirm comparable sizes of the dust and UV continuum do not come from observational uncertainties.
The extended dust size may be due to (i) clumpy gas distribution makes UV photons possible to radiate molecular clouds around the outskirts of galaxies or (ii) rest-frame UV observations do not trace global star formation activities but massive star-forming clumps within galaxies.

\item The CRISTAL galaxies follow a similar trend as LTGs or LBGs in the stellar mass-size plane (Figure \ref{fig:masssize}).
They are distributed in a different regime compared with more dusty galaxies that follow the trend of ETGs.  
Our result suggests typical star-forming galaxies with $M_{\ast}\sim10^{10}\,M_{\odot}$ are in a growing-up phase of disks through global star formation.

\item An expansion of the disks in $z\sim5$ normal SFGs along with the star-forming sequence in the mass-size plane are completely consistent with the large stellar disks of massive galaxies at $z\sim2$. 
An expansion of the stellar disks from $z\sim5$ to $z\sim2$ and the subsequent compaction events naturally explain morphological transformation and are consistent with previous studies examining massive galaxies at $z\sim2$.

\end{enumerate}

Our results emphasize the importance of considering the spatial extent of both the dust-unobscured and obscured star-formation activities are important when studying their morphological transformation, especially given more than $\sim$50\% of the star-formation activities is obscured by the dust at the stellar mass range of $M_{\ast}>10^{10}\,M_{\odot}$.
Since the spatial extent of the dust-obscured star formation is almost comparable with that of the dust-unobscured one, normal star-forming galaxies at $z\sim5$ are not making their sizes compact, but expanding their stellar disks.
Inner structures of the typical star-forming galaxies at $z\sim5$, at the end of the epoch of reionization, is an important question to be addressed with future ALMA and {\it JWST} observations, which will provide the more secure spatial extent of dust-obscured star-forming regions or other components such as pre-existing stellar disks or ionized gas.
Also, ALMA has the capability of high-frequency observations such as band-8/9/10, which enables us to constrain the FIR SED of each galaxy and bring us to a clearer picture of the obscured star formation.

\begin{acknowledgements}
This paper makes use of the following ALMA data: ADS/JAO.ALMA\#2017.1.00428.S, \#2018.1.01359.S, \#2018.1.01605.S, \#2019.1.01075.S, \#2019.1.00226.S, and \#2021.1.00280.L. 
ALMA is a partnership of ESO (representing its member states), NSF (USA), and NINS (Japan), together with NRC (Canada), MOST and ASIAA (Taiwan), and KASI (Republic of Korea), in cooperation with the Republic of Chile. The Joint ALMA Observatory is operated by ESO, AUI/NRAO, and NAOJ.
This work was supported by JSPS KAKENHI grant No. 23K03466.
I.M. is financially supported by Grants-in-Aid for Japan
Society for the Promotion of Science (JSPS) Fellows
(KAKENHI Number 22KJ0821).
R.H.-C. thanks to the Max Planck Society for support under the Partner Group project "The Baryon Cycle in Galaxies" between the Max Planck for Extraterrestrial Physics and the Universidad de Concepción. 
R.H-C. also gratefully acknowledges financial support from Millenium Nucleus NCN19058 (TITANs), and ANID BASAL projects ACE210002 and FB210003.
M.A. acknowledges support from FONDECYT grant 1211951. 
M.A. and R.J.A. acknowledge support from ANID BASAL project grant FB210003. 
R.J.A. was supported by FONDECYT grant number 123171.
R.B. acknowledges support from an STFC Ernest Rutherford Fellowship [grant number ST/T003596/1].
R.D. is supported by the Australian Research Council Centre of Excellence for All Sky Astrophysics in 3 Dimensions (ASTRO 3D), through project number CE170100013. 
A.F. acknowledges support from the ERC Advanced Grant INTERSTELLAR H2020/740120.
T.N. acknowledges support from the Deutsche Forschungsgemeinschaft (DFG, German Research Foundation) under Germany’s Excellence Strategy - EXC-2094 - 390783311 from the DFG Cluster of Excellence ``ORIGINS''.
M.R. acknowledges support from project PID2020-114414GB-100, financed by MCIN/AEI/10.13039/501100011033. 
H.{\"U}. gratefully acknowledges support by the Isaac Newton Trust and by the Kavli Foundation through a Newton-Kavli Junior Fellowship.
Data analysis was in part carried out on the Multi-wavelength Data Analysis System operated by the Astronomy Data Center (ADC), National Astronomical Observatory of Japan.
\end{acknowledgements}

\appendix

\section{CIGALE parameters used to estimate global properties}
We use all available broad- (BBs) and medium bands (MBs) in optical to near-infrared (NIR) for the SED fitting.
In the COSMOS2015 catalog, we use 10 broad bands ($u^{\ast}$, $B$, $V$, $r^+$, $i^+$, $z^{++}$, $Y$, $J$, $H$, $K_s$) and 12 medium bands on Ground-based telescopes, and 4 {\it Spitzer} bands that achieve the best sensitivities among the bands covering similar wavelength. 
We take 4 broad bands ($U$, $B$, $R$, $K_s$) and 23 medium bands on Ground-based telescopes, and 10 {\it HST} and 4 {\it Spitzer} bands on space telescopes from the ASTRODEEP catalog.
We extract the total fluxes of the counterparts of the CRISTAL galaxies from the corresponding catalogs.
Table \ref{tab:tab2} shows parameters used in SED fitting to derive stellar masses and SFRs (see Section \ref{subsec:globsed}).

\begin{table}
\centering
\caption{Parameters for SED fitting}
%\tablewidth{0pt}
\begin{tabular}{lc}
\hline
Parameters & Range \\
\hline
\hline
\multicolumn{2}{c}{{\bf SFH: Delayed-$\tau$ ({\sc delayed})}} \\
\hline
$\tau_{\rm main}$ [Myr] & 100, 300, 500, 1000, 2000, 5000 \\
${\rm Age}_{\rm main}$ [Myr] & 50\,Myr steps in [50: 1000] \\
$\tau_{\rm burst}$ [Myr] & 10, 50, 100\\
${\rm Age}_{\rm burst}$ [Myr] & 10, 20, 50 \\
${\rm f}_{\rm burst}$ [Myr] & 0.0, 0.05, 0.1, 0.2 \\
\hline
\multicolumn{2}{c}{{\bf SSP: BC03}} \\
\hline
IMF & Chabrier \\
$Z$ [$Z_{\odot}$] & 0.2, 1\\
\hline
\multicolumn{2}{c}{{\bf Nebular emission: CLOUDY}} \\
\hline
$\log_{10} U$ & -3.0 \\
line width [\kms] & 300\\
\hline
\multicolumn{2}{c}{{\bf Dust attenuation: Calzetti00}}\\
\hline
E(B-V) & 0.1 steps in [0.0: 2.0] \\
E(B-V) factor & 0.44\\
$R_{\rm v}$ & 3.1 \\
\hline
\end{tabular}
\label{tab:tab2}
\end{table}

\section{comparison of ${\rm SFR}_{\rm SED}$ and ${\rm SFR}_{\rm UV}+{\rm SFR}_{\rm IR}$}
\begin{figure}[htbp]
\centering
\includegraphics[width=9cm]{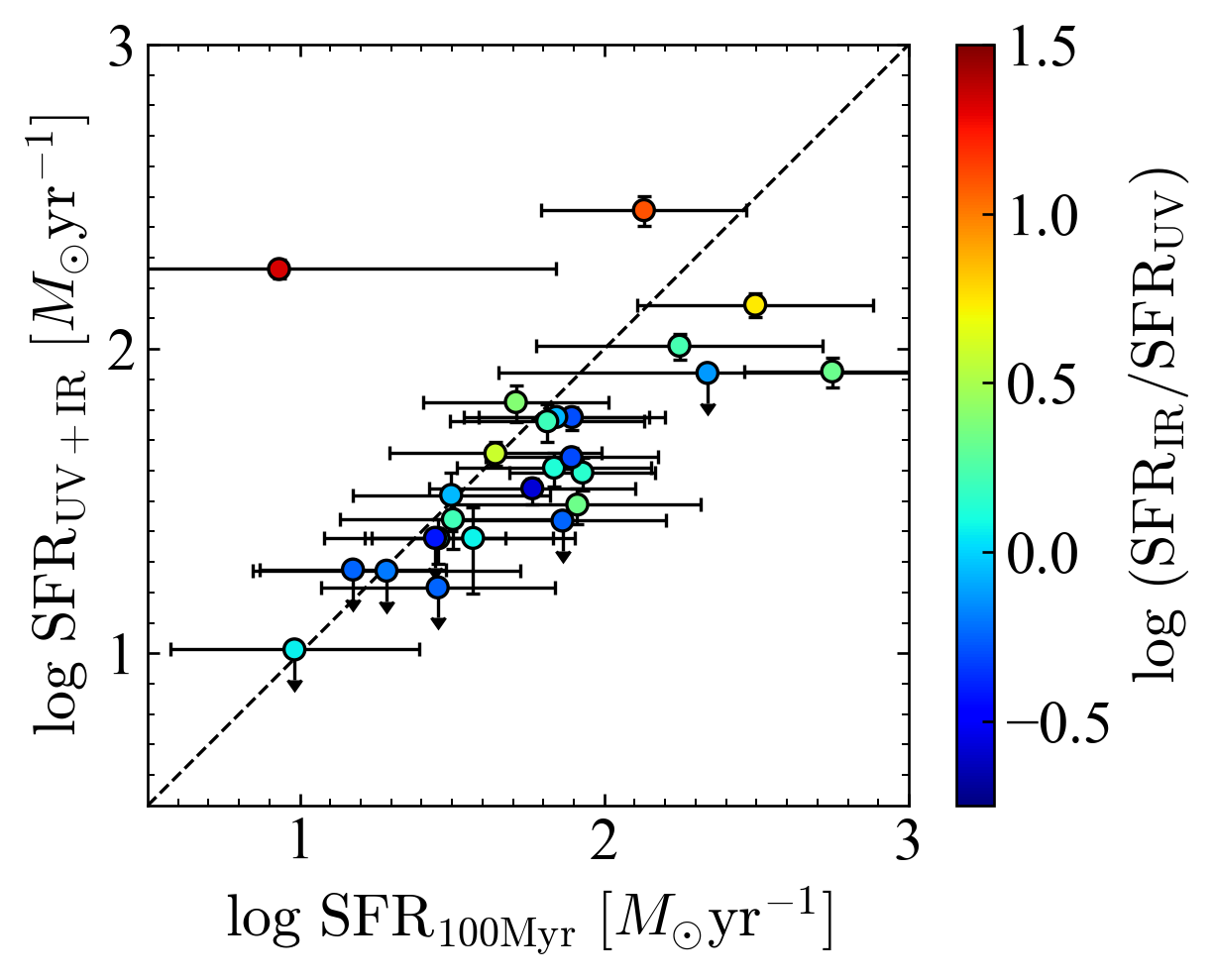}
%\captionsetup{labelformat=empty,labelsep=none}
\caption{${\rm SFR}_{\rm UV}+{\rm SFR}_{\rm IR}$ vs ${\rm SFR}_{\rm SED}$. The black dashed line represents a one-to-one relationship. ${\rm SFR}_{\rm SED}$ is broadly consistent with ${\rm SFR}_{\rm UV}+{\rm SFR}_{\rm IR}$, except for CRISTAL-24.}
\label{fig:sfruvir}
\end{figure}

In Figure \ref{fig:sfruvir}, we compare ${\rm SFR}_{\rm UV}+{\rm SFR}_{\rm IR}$ and ${\rm SFR}_{\rm SED}$.
${\rm SFR}_{\rm UV}+{\rm SFR}_{\rm IR}$ is calculated by summing up ${\rm SFR}_{\rm UV}$ from best-fit SED and ${\rm SFR}_{\rm IR}$ from the rest-frame $158\,\mu{\rm m}$ dust continuum flux and ${\rm SFR}_{\rm SED}$ is computed by averaging the SFH over $100\,{\rm Myr}$.
${\rm SFR}_{\rm SED}$ is broadly consistent with ${\rm SFR}_{\rm UV}+{\rm SFR}_{\rm IR}$.
CRISTAL-24 show $\sim1.4\,{\rm dex}$ smaller ${\rm SFR}_{\rm SED}$ than ${\rm SFR}_{\rm UV}+{\rm SFR}_{\rm IR}$.
This difference is due to the degeneracy of the dust extinction and stellar age and makes CRISTAL-24 distribute $\sim1.2\,{\rm dex}$ lower than the main sequence.

%
%
%
% Figure 13
%
%
%

\section{Validation of the dust size measurement}
As shown in Table \ref{tab:tab1}, about half of the CRISTAL galaxies range $4.4<{\rm S/N}<5.5$ in the dust continuum images. 
These S/Ns might translate to significant uncertainties in our flux measurements but also to systematical errors in the sizes. 
To evaluate uncertainties of the size measurement through visibility fitting and confirm the validity of the size measurement, we conducted a Monte Carlo simulation and a stacking analysis toward fainter CRISTAL galaxies.

\subsection{Monte-Carlo simulation}\label{subsubsec:MCsim}
Because the measured sizes might be affected by noise fluctuations, we need to know the dependence of the size measurement uncertainties on the source S/N.

In the following Monte-Carlo simulation, we use the visibilities of six galaxies (CRISTAL-01b, 07ab, 09, 10, 11a, and 19) that have relatively low S/Ns among the CRISTAL sample ($4.4\leq{\rm S/N}\leq5.5$).
This selection was made because galaxies with smaller S/N are anticipated to be more affected by noise fluctuations in size measurements.
%because the more severe effect of noise fluctuations is expected in the size measurements for the galaxies with smaller S/Ns.
We take different array configurations used in each CRISTAL galaxy into account as the visibilities of the six galaxies encompass various representative $uv$ coverages among those obtained in our observations.
We create $\sim1500$ ($\sim300$ in each field) artificial circular Gaussian sources with a log uniform distribution of total flux within a range of 0.025--1.25 mJy and effective radius ranging 0.025--$1.5''$.
After subtracting observed galaxies with the CASA task {\sc tclean}, we injected the generated mock sources in the visibilities at a random position and calculated source S/N in the same manner as referred to in Sec \ref{subsec:dustdetect}.
When the mock source is detected at $4.4<{\rm S/N}$ (lowest S/N among observed galaxies in dust continuum), we fit a circular Gaussian model by setting an initial estimation of the central position to the artificial source center using {\sc UVMultifit}.

\begin{figure*}[htbp]
\begin{center}
%\epsscale{1.15}
\includegraphics[width=16cm]{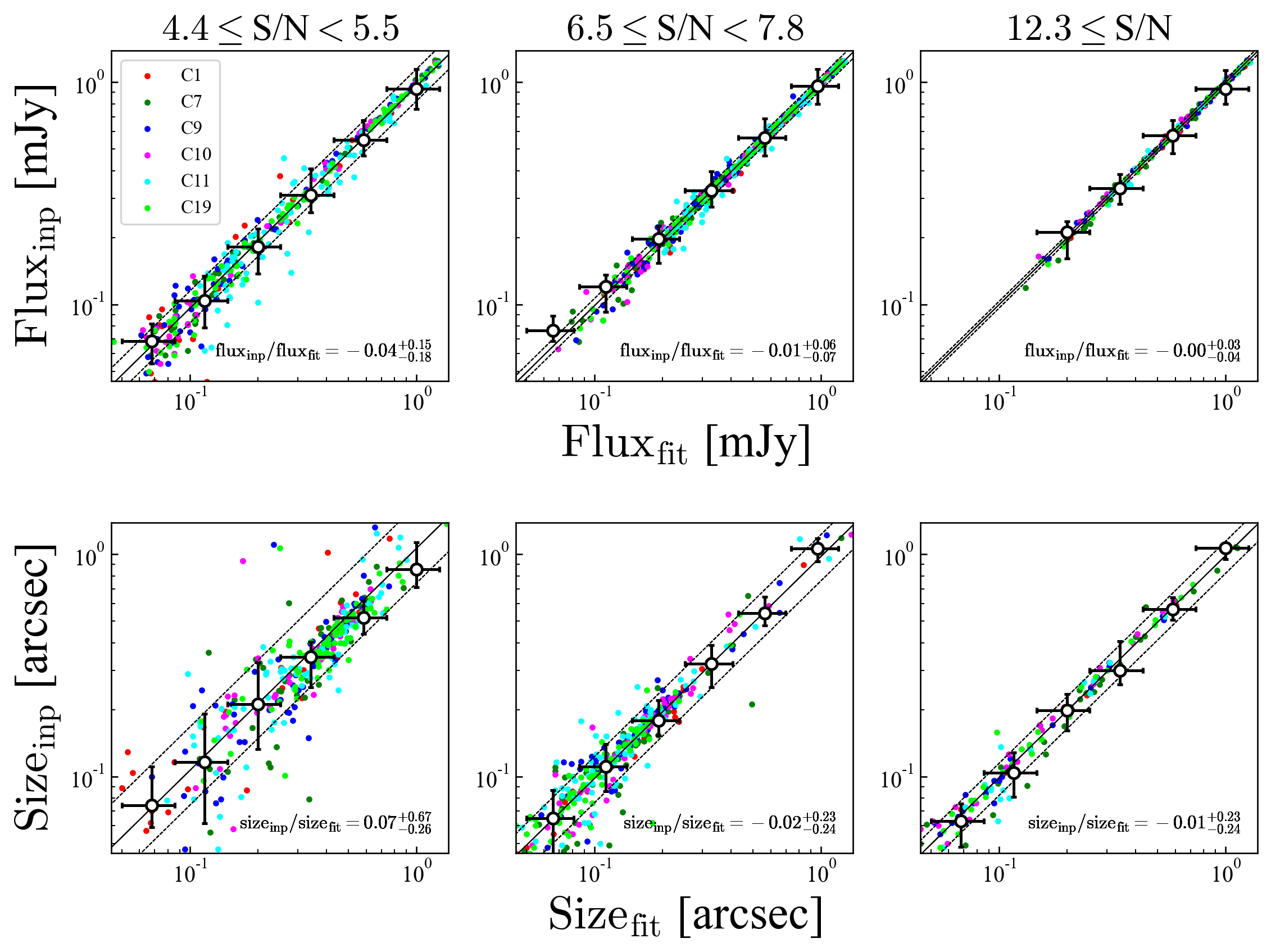}
\caption{Comparison of the flux density and FWHM between fitting results and input models. The left, middle, and right panels show the results for sources detected at ${\rm S/N}=4.4$--5.5, ${\rm S/N}=6.5$--7.8 and ${\rm S/N}\geq12.3$, respectively. Different colors indicate different visibilities used in the simulation, corresponding to the different array configurations. White circles and error bars show $50\pm34\%$ percentiles in the bins of output values. Black solid and dashed lines correspond to the averages and standard deviations of the fitting results and output models, which are also shown in the bottom right of each panel.}
\label{fig:MCresults}
\end{center}
\end{figure*}

In Figure \ref{fig:MCresults}, we plot the result of the Monte-Carlo simulation, with an average and standard deviation of the ratio between input and output values.
As shown in Figure \ref{fig:MCresults}, independent of the S/N of the source, we do not find any systematic offsets between intrinsic sizes and resulting measured sizes (${\rm size}_{\rm inp}/{\rm size}_{\rm fit}\sim0$).
This result demonstrates the estimated sizes of these galaxies are on average reliable.
However, careful treatment of individual sizes of the sources with $4.4\leq{\rm S/N}\leq5.5$ is needed as the output sizes of these sources show large scatters from input sizes (${\rm size}_{\rm inp}/{\rm size}_{\rm fit}=0.07_{-0.26}^{+0.67}$) especially in the case that output sizes are less than $0.4''$ (${\rm size}_{\rm inp}/{\rm size}_{\rm fit}=0.09_{-0.32}^{+0.83}$).
To obtain a better estimation of the size of these low-S/N sources, we additionally carry out a stacking analysis in the following section.

%
%
%
% Figure 11
%
%
%

\subsection{Stacked size}\label{subsec:stack}
We now perform stacking analysis to check whether there are systematic biases in our size estimates. 
If galaxies included in stacking are intrinsically compact (extended) but appear as extended (compact) due to noise effects, stacked sizes are expected to become smaller (larger) than a weighted average of the individual sizes.
We perform the stacking analysis, especially for six of eight CRISTAL galaxies with $4.4\leq{\rm S/N}\leq5.5$ (CRISTAL-01b, 07ab, 09, 10, 11a, and 19), to evaluate the potential size misestimations. 
We did not use the other two galaxies (CRISTAL-03 and 21) because CRISTAL-03 is not resolved and CRISTAL-21 has extremely extended sizes therefore they may affect stacked sizes strongly.

%Before we proceed with stacking on the observed CRISTAL galaxies, we conduct stacking on artificial sources, for which we already know the intrinsic sizes, to validate whether the size estimates are proper.
%we confirm whether the stacking brings us proper size estimation with artificial sources because we know the intrinsic sizes of the sources.
%We produce the artificial sources having $4.4\leq{\rm S/N}\leq5.5$ made during the Monte-Carlo simulation above. 
%We align phase centers to the measured central position and set the phase center to a common coordinate of $\alpha$ = 00h 00m 00s; $\delta$ = 00d 00m 00s (J2000) with CASA task {\sc fixplanet}.
%We concatenate all visibilities into a single measurement set with the {\sc concat} task.
%Finally, we measured the sizes with {\sc UVMultifit}.
%As a result, when we stack the sources whose individual sizes are overestimated by $\sim20\%$ with respect to intrinsic sizes, the stacked size becomes $\sim10\%$ more compact than the average of the individually measured sizes.
%In other words, the stacked size is close to the average of the intrinsic sizes.
%To avoid the stacked size being dominated by apparently bright sources, we also try scaling the amplitude included in the measurement set to 10~$\mu$Jy using the CASA task {\sc gencal} and {\sc applycal} and assigning weights inversely proportional to the square of the total flux through the {\sc visweightscale} parameters when we concatenating all visibilities with the {\sc concat} task.

We align phase centers to the measured central position and set the phase center to a common coordinate of $\alpha$ = 00h 00m 00s; $\delta$ = 00d 00m 00s (J2000) with CASA task {\sc fixplanet}.
We concatenate all visibilities into a single measurement set with the {\sc concat} task.
Finally, we measured the sizes with {\sc UVMultifit}.

%Then, we apply the same stacking procedure on six of eight CRISTAL galaxies with $4.4\leq{\rm S/N}\leq5.5$ (CRISTAL-01b, 7, 9, 10, 11, and 19).
The estimated average size is $0.209\pm0.061''$, which is consistent with the average of the individually measured sizes ($0.237''$) within $1\sigma$ uncertainty.
%As these results are consistent with that from the mock sources, the intrinsic sizes of the six CRISTAL galaxies listed above may be overestimated by $\sim20\%$.
%While $\sim20\%$ uncertainties do not significantly affect the conclusion, we also plot the stacked size in Figure \ref{fig:masssize}.
Therefore we conclude that the individual measurements of the six CRISTAL galaxies listed above are reliable.
While the stacked measurements do not significantly affect the conclusion, we also plot the stacked size in Figure \ref{fig:masssize}.

\section{S{\'e}rsic index-dependence of the UV sizes}
\begin{figure}[htbp]
\centering
\includegraphics[width=9cm]{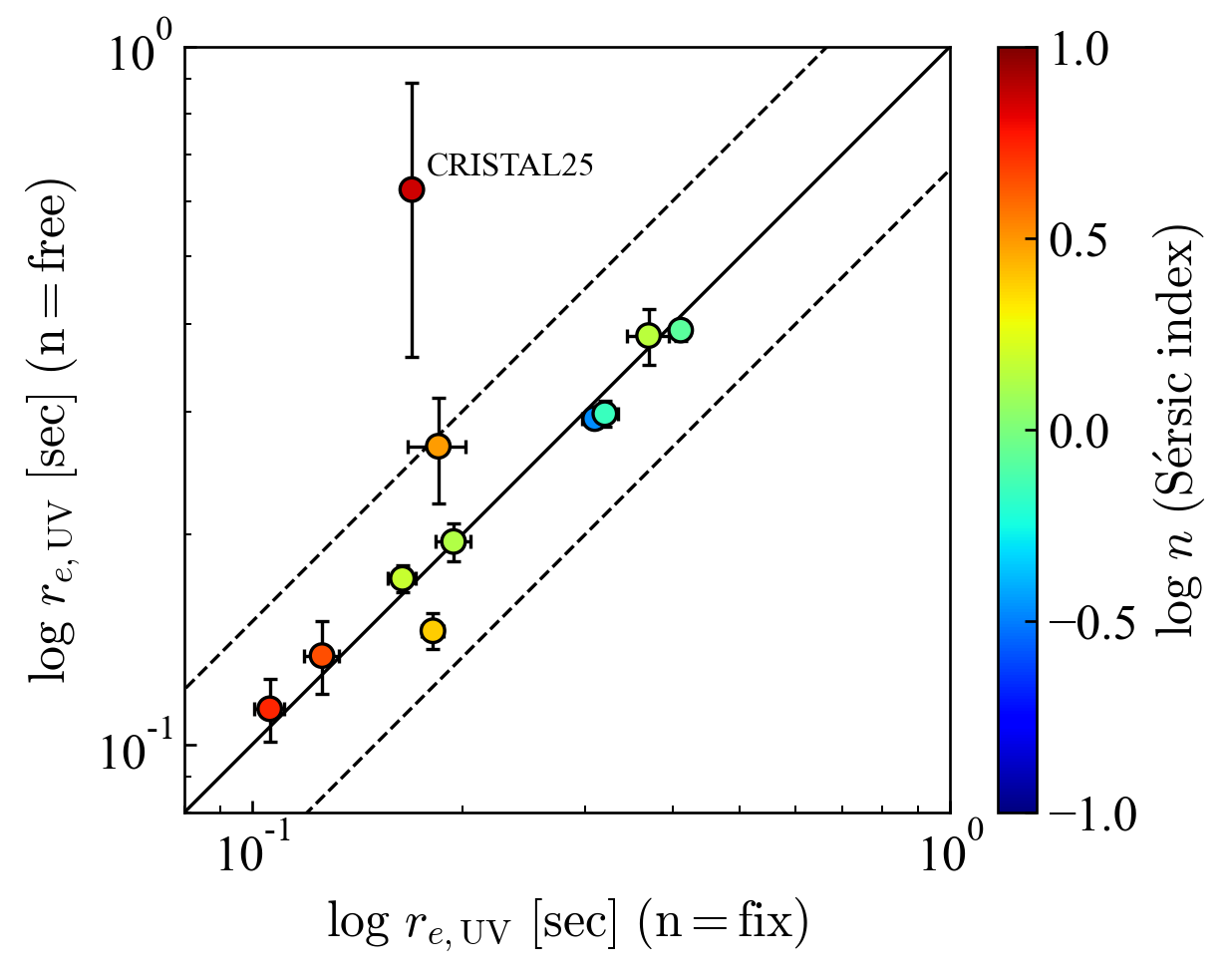}
%\captionsetup{labelformat=empty,labelsep=none}
\caption{UV size comparison between the case with the S{\'e}rsic index fixed ($n=1$) and with the S{\'e}rsic index free. Black solid and dashed lines indicate one-to-one relation and $\pm0.3{\rm dex}$. The sizes are almost consistent, except for CRISTAL-25 showing a large S{\'e}rsic index ($n=7.3\pm1.7$).}
\label{fig:sizecomp}
\end{figure}

In Figure \ref{fig:sizecomp}, we compare the derived UV sizes under the assumption of S{\'e}rsic index $n=1$ and $n={\rm free}$.
Overall, there is no systematic difference between the two sizes.
CRISTAL-25 shows a larger size in the case of $n={\rm free}$ because the best-fit S{\'e}rsic index is larger than unity ($n=7.3\pm1.7$).
We note that this size difference of CRISTAL-25 does not affect the conclusion.

%
%
%
% Figure 11
%
%
%

\section{fitting results of ALMA and HST}
\begin{figure*}[htbp]
%%%\begin{center}
\vspace{1cm}
%\epsscale{1.0}
\includegraphics[height=18cm]{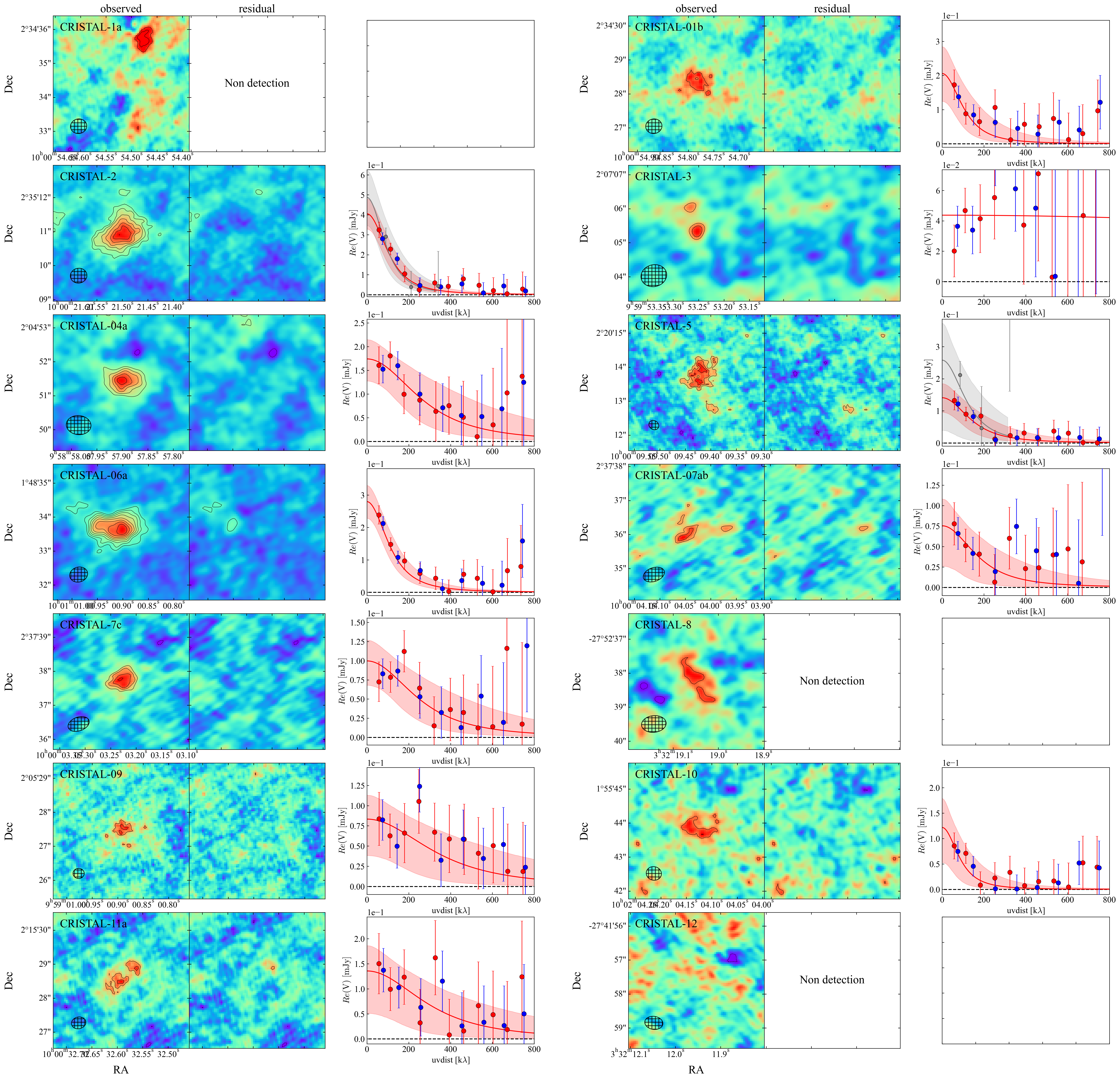}
\caption{Fitting results of the ALMA images. The left and middle panels show the natural-weighted observed and residual images respectively. The contours indicate every 1$\sigma$ from $\pm3\sigma$ to 10$\sigma$ and every 3$\sigma$ from 10$\sigma$. The right panels show the visibility real part as a function of $uv$-distance from a central position of the fitting in two different binning scales (red and blue). Red solid lines and shaded regions show best-fit models and 1$\sigma$ uncertainties. We also show sizes without the CRISTAL dataset when the fittings have converged. (gray points and solid lines).}
\label{fig:fittingALMAnoT}
%%%\end{center}
\end{figure*}

\begin{figure*}[htbp]
%\begin{center}
%\epsscale{1.15}
\includegraphics[height=18cm]{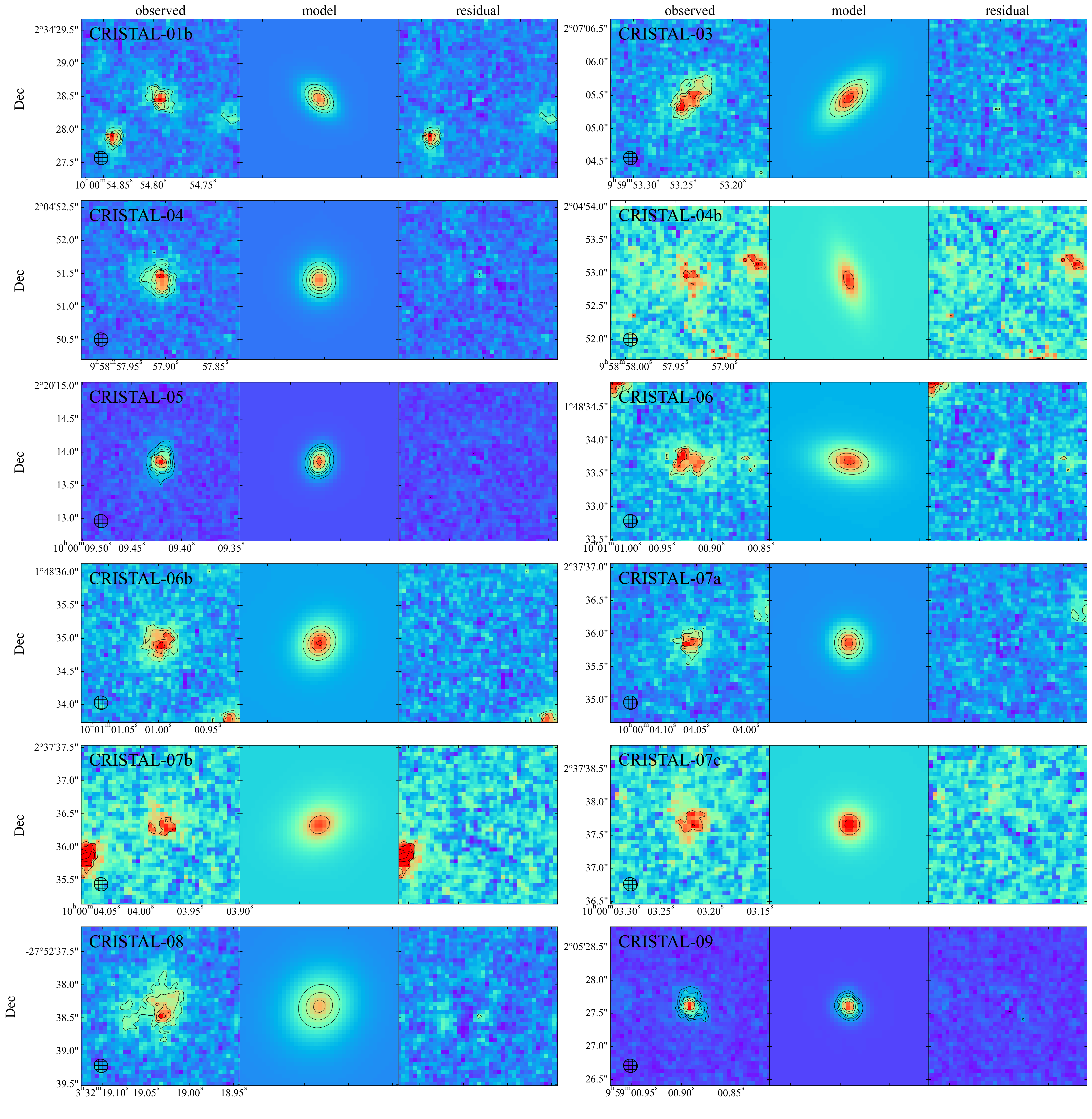}
\caption{The fitting results of the {\it HST}/F160W images. Observed, model, and residual images are shown from left to right. The contours indicate every 2$\sigma$ from $\pm4\sigma$.} %Circularized effective radii are shown in the top left of the model columns.}
\label{fig:fittingHST}
%\end{center}
\end{figure*}

Figure \ref{fig:fittingALMAnoT} shows the results of the dust continuum size measurements in the visibility domain. The left and middle panels show the observed and residual images respectively. The observed and model images are natural-weighted in Figure \ref{fig:fittingALMAnoT}.
%and $0.5''$ $uv$-Tapers are additionally adopted in \ref{fig:fittingALMAT05}. 
The right panels show visibility amplitudes as a function of $uv$-distance from the central position of the fitting in two different binning scales (red and magenta). Red solid lines and shaded regions show best-fit models and uncertainties. To illustrate how the sizes are measured without the CRISTAL dataset, we add these results in gray points and solid lines. 
%Note that almost all of the CRISTAL galaxies are not detected in the dust continuum without the CRISTAL dataset.
Figure \ref{fig:fittingHST} shows results of the rest-frame UV size measurements in {\it HST}/F160W.  

%
%
%
% Figure A1
%
%
%
%
%
%
%
%
%

%
%
%
% Figure A1
%
%
%
\begin{figure*}[htbp]
%%%\begin{center}
\vspace{1cm}
%\epsscale{1.0}
\includegraphics[height=15cm]{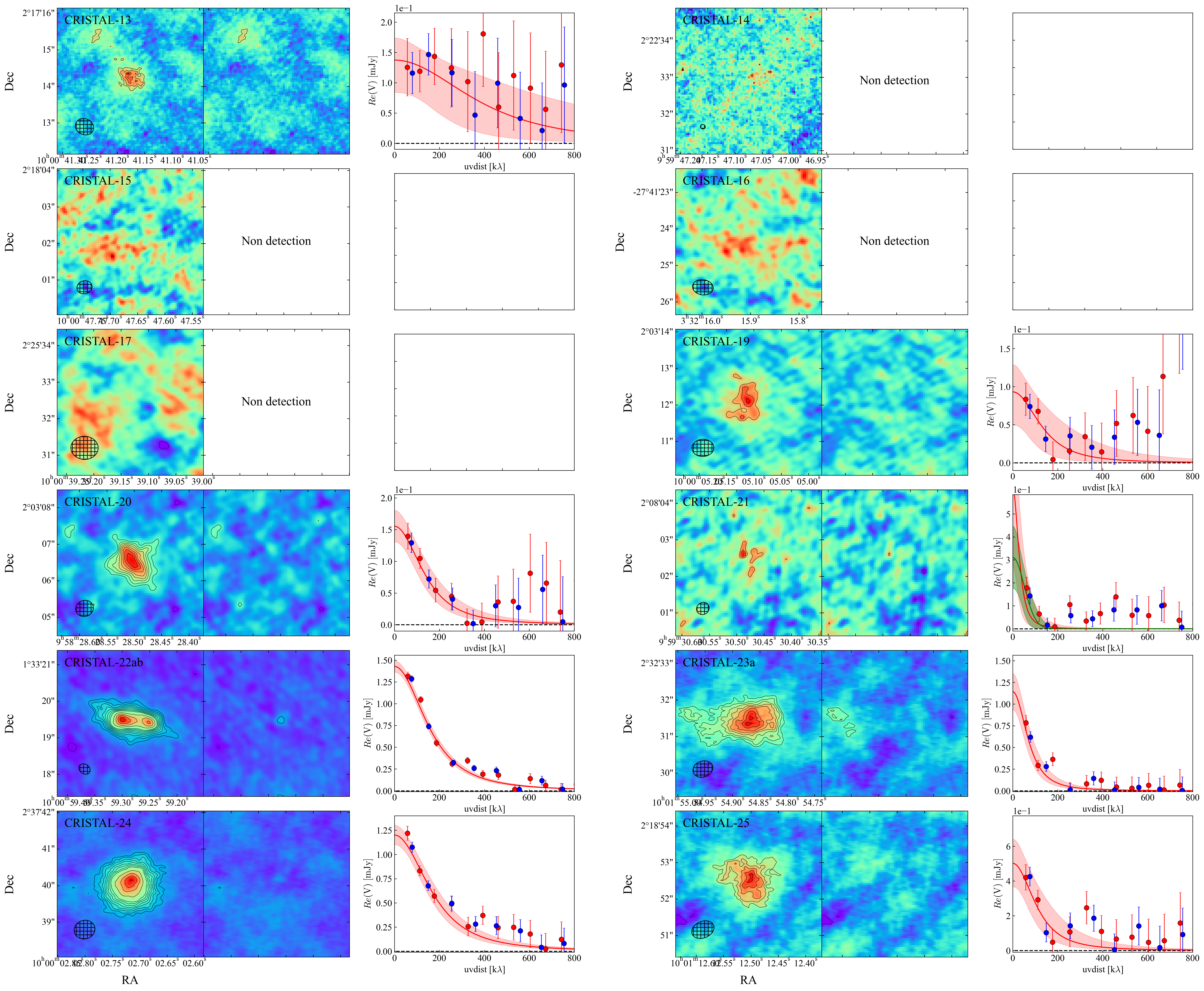}
\caption{Continue from Figure \ref{fig:fittingALMAnoT2}. For CRISTAL-21, we also show the best-fit profile assuming Gaussian to illustrate how the measurement is uncertain due to the extrapolation toward the uv distance of $<50\,{\rm k}\lambda$}
%\caption{Continue from Figure \ref{fig:fittingALMAnoT2}. For CRISTAL-21, we also show the fitting result assuming the Gaussian profile in the green-shaded region.}
\label{fig:fittingALMAnoT2}
%%%\end{center}
\end{figure*}
%
%
%
%
%
%

%
%
%
% Figure A2
%
%
%
%\begin{figure*}[htbp]
%%%\begin{center}
%\vspace{1cm}
%%\epsscale{1.0}
%\includegraphics[height=22cm, bb=0 0 1800 2550, trim=0 1 0 0cm]
%{fitting_results_Taper05sec.png}
%\caption{The same as Figure \ref{fig:fittingALMAnoT}, but the $0.5''$ $uv$-Tapers are additinaly adopted in imaging.}
%\label{fig:fittingALMAT05}
%\end{center}
%\end{figure*}
%
%
%
%
%
%

%
%
%
% Figure A3
%
%
%
%
%
%
%
%
%

%
%
%
% Figure A3
%
%
%
\begin{figure*}[htbp]
%\begin{center}
%\epsscale{1.15}
\includegraphics[height=18cm]{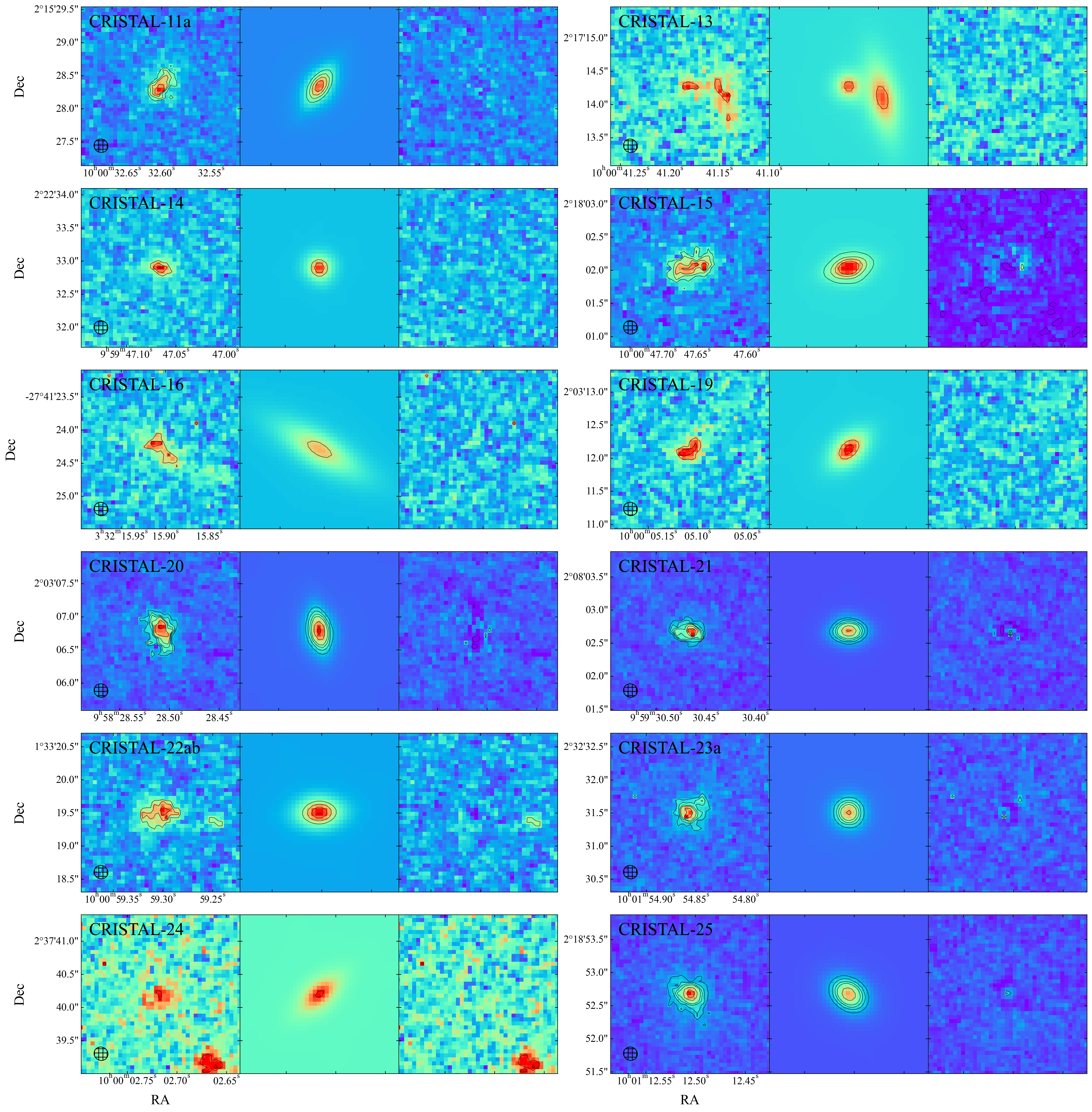}
\caption{(continue)}
\label{fig:fittingHST2}
%\end{center}
\end{figure*}

\clearpage
\bibliography{ref.bib}{}
\bibliographystyle{aa.bst}

\end{document}